\documentclass[a4paper,11pt]{article}
\usepackage[left=1in,right=1in,top=1in,bottom=1in]{geometry}
\usepackage[english]{babel}
\usepackage{physics}
\usepackage{authblk}
\usepackage[autostyle]{csquotes}
\usepackage{amssymb}
\usepackage{amsfonts}
\usepackage{graphicx}
\usepackage{epsfig}
\usepackage[stable]{footmisc}
\usepackage{appendix}
\usepackage{dsfont}
\usepackage{lmodern}
\usepackage{simpler-wick}
\usepackage{bbm}
\usepackage{bm}
\usepackage{subfig}
\usepackage{cancel}
\usepackage[nottoc, notlof, notlot]{tocbibind}
\usepackage[colorlinks=true, breaklinks, linktocpage=true]{hyperref}
\usepackage{array}
\usepackage{longtable}
\makeatletter
\g@addto@macro\bfseries{\boldmath}
\makeatother

\newcolumntype{P}[1]{>{\centering\arraybackslash}m{#1}}

\newcommand{\half}{\frac{1}{2}}

\newcommand{\T}{{\Theta}} 
\newcommand{\ra}{\rangle}
\newcommand{\la}{\langle}

\newcommand{\cN}{\mathcal{N}}
\newcommand{\del}{\partial}
\newcommand{\Rb}{\bar {R}}

\def\be{\begin{equation}}
\def\ee{\end{equation}}
\def\bs{\begin{split}}
\def\es{\end{split}}
\def\bea{\begin{eqnarray}}
\def\eea{\end{eqnarray}}
\def\bs{\begin{split}}
\def\es{\end{split}}
\def\ie{\begin{equation}\begin{aligned}}
\def\fe{\end{aligned}\end{equation}}

\title{{\bf Superconformal invariants and spinning correlators in $3d$  ${\cal N}=2$ SCFTs}}
\author{{Aditya Jain and Amin A. Nizami}
\thanks{email: adityaj2807@gmail.com, aan27cam@gmail.com}}
\affil{\small Department of Physics, Ashoka University, \\
Rajiv Gandhi Education City, NCR, India, 131029}
\date{} 

\begin{document}
\maketitle

{\abstract  We construct superconformal invariants in  superspace which are used to build 3-point correlators of spinning operators in general $\mathcal{N}=2$ superconformal field theories in three dimensions. Our systematic analysis includes various relations between these invariants and provides a minimal set of parity-even and parity-odd invariants which is further used to construct general 3-point functions in any 3d $\mathcal{N}=2$ SCFT. For conserved (super)currents, we explicitly compute various 3-point functions using Wick contractions in the free field case, and express them in terms of the constructed parity-even invariants. We give evidence through examples for the claim that the 3-point function of conserved currents generally comprises of two parts - a parity-even piece coming from the free theory, and a parity-odd piece.}
%\keywords{Conformal Field Theory, Supersymmetry, Superspace}
%\preprint{}
\vfill
\pagebreak

%\makeatletter
%\pdfstringdefDisableCommands{\let\HyPsd@CatcodeWarning\@gobble}
%\makeatother

%\makeatletter
%\gdef\@fpheader{}
%\makeatother

\tableofcontents

\section{Introduction} 
Superconformal invariance is the largest spacetime symmetry possible in a local relativistic quantum field theory. Consequently, it is extremely constraining. Conformal algebras exist in all space-time dimensions but the supersymmetry algebra can be extended to a closed superconformal algebra only for $d \le 6$ \cite{Nahm:1977tg}. Also, as is well known, two and three point functions of operators are completely fixed by (super)conformal symmetry alone \cite{Polyakov:1970xd, Schreier:1971um}. For the case of 3-point functions of spinning operators, the form is still exactly fixed, but a number of tensor structures are possible \cite{Sotkov:1976xe, Osborn:1993cr}, each coming with an OPE coefficient encoding the dynamics of the theory. Since the number of these tensor structures, as well as the complexity in the form of each, increases rapidly with increasing spin of the correlator, it is useful to have an efficient and tractable way of keeping track of the tensorial nature of such correlators. For non-supersymmetric theories, one such approach was provided by \cite{Giombi:2011rz} (see also \cite{Costa:2011mg} which utilises the embedding space formalism).

In this paper, we will study the constraints of $\cN=2$ superconformal invariance on 3-point correlators of spinning primary operators in three dimensions. We will extensively use the superspace formalism \cite{Salam:1974jj} and work throughout with correlators of superfields. In particular, we will construct a multitude of 3-point superconformal invariants and express 3-point correlators as multinomials in these invariants. Using the polarisation spinor formalism \cite{Giombi:2011rz} and these invariants, we will find that the proliferating complexity of tensor structures with increasing spin becomes tractable, and one only needs to deal with multinomials built out of a handful of these invariants to encode the tensor structure of general spinning correlators. Furthermore, we will study the constraints that arise when one or more operators are conserved currents. 

Related studies of various aspects of 3-point correlators in 3d SCFTs include \cite{Park1999cw, Nizami:2013tpa, Buchbinder:2015qsa, Kuzenko:2015lfa, Buchbinder:2015wia, Kuzenko:2016cmf, Buchbinder:2021qlb, Buchbinder:2021gwu}.  %in three dimensions and \cite{Park:1997bq, Osborn:1998qu, Buchbinder:2022msd, Buchbinder:2022cqp} in four dimensions. 
For an overview of the general constraints of superconformal invariance in various dimensions, see \cite{Minwalla:1997ka, Dolan:2001tt}.

This paper is structured as follows. We begin by reviewing the essential details of 3d superspace in Section \ref{Scov}. Here we also construct the superconformal covariant structures which are used in later sections to build 3-point function invariants. In Sections \ref{PE}, \ref{PO} we construct the primary 3-point invariants, both parity-even and parity-odd, out of which all 3-point functions of spinning operators are built.  The relations that exist between these invariants are also discussed in these sections. In Section \ref{3pt} we give examples of various 3-point correlators expressed as a linear combination of linearly independent structures built out of the invariants. We move on to discuss correlators of conserved currents in section \ref{CC}. We first fix the form of the free-theory 3-point correlators in terms of parity-even invariants. Subsequently, using the appropriate conservation (multiplet shortening) condition we give evidence through examples that the general 3-point conserved current correlator in an $\cN =2$ 3d SCFT
is the sum of a parity-even (free theory) part and a parity odd part, with two undetermined OPE coefficients. We conclude with a discussion and future directions in Section \ref{dis}.

Some technical details are relegated to the appendices.
Appendix \ref{conv} contains details of the various conventions used and lists various useful relations.
Appendix \ref{rel} lists various relations where invariants constructed out of the covariant structures are all expressed in terms of our minimal set of invariants.
Appendix \ref{FT} details, through two examples, the computation of 3-point conserved current correlators in a free theory.

\section{Superconformal covariant structures in superspace}\label{Scov}

In this section, we provide an overview of the required basic features of superspace and construct the various covariant structures out of which 3-point functions are built. This will help set the stage for constructing 3-point invariants and correlators in the following sections. 

A foundational work on 3d SCFTs in superspace is \cite{Park1999cw} to which the reader is referred to for additional details. %- see also \cite{}. 
We have included the necessary details in this section to make the paper self-contained for the most part.
Note however that our conventions are different - they are those of \cite{Nizami:2013tpa} and are summarised in Appendix \ref{conv}. We will also use {\it polarisation spinors} to encode spin, in a manner similar to \cite{Giombi:2011rz}, as explained in more detail below. 
% Our techniques will thus be an amalgamation of those of \cite{GPY} and \cite{Park}.

Superspace is parametrised by coordinates $(x^{\mu}_{i},\theta^{a}_{i\alpha}) $ where $x$ denotes the usual bosonic coordinate, $\theta$ denotes the fermionic (grassmanian) coordinate, $\mu=0,1,2$ is a Lorentz index, $\alpha=1,2$ is a spinor index, $a=1,2,....,\cN$ is the R-symmetry index and $i=1,2,3$ will label a particular superspace point.\footnote{Throughout the paper, repeated spinor and R-symmetry indices ($\alpha, a$) will be summed over whereas repeated superspace point indices $i, j$ etc. will {\it never} be summed unless explicitly mentioned.} We will work with Majorana spinors, the R-symmetry group being $SO(2)$ for the $\cN=2$ case we will be focus on. Since $\delta_{ab}$ is the only invariant tensor for $SO(2)$, all R-symmetry singlets will be built by contracting R-symmetry charged objects with $\delta_{ab}$'s.

Following \cite{Giombi:2011rz} we will use spinor notation to write an $SO(2,1)$ vector as a bi-spinor - $X_{\alpha}^ {\,\,\beta}=x^{\mu} (\gamma_{\mu})_{\alpha}^{\,\,\beta} $, where $\gamma_{\mu}$ are the 3d `Pauli' matrices. Their form along with various other conventions used in the paper  are given in Appendix  \ref{conv}.

The action of (super)inversion on the superspace coordinates is given by 
\begin{equation} \begin{split} \label{inver}
I(x^\mu)&= \frac{x^\mu}{x^2 + \frac{(\theta^a\theta^a)^2}{16}}\,\,\,, \\
% I(\theta^a_\alpha)&= \\
\end{split}
\end{equation}
and 
\begin{equation}
\begin{gathered}
I(\theta^a_{\alpha}) = (X_{+}^{-1}\theta^a)_{\alpha},\\
I(\theta^{a\beta})=-(\theta^a X_{-}^{-1})^{\beta},
\end{gathered}
\end{equation}
where we have defined
\begin{equation}\label{pmdef}
X_{\pm}=X\pm \frac{i}{4}(\theta^a\theta^a)\mathds{1},\end{equation}
which transforms under super-inversion as
\[
I(X_{\pm}) =X_{\pm}^{-1}.\]

Now the following two structures, constructed out of 2 superspace points, are annihilated by the supersymmetry generators:
\begin{gather} \label{tx}
\tilde{x}_{12}^{\mu}=x_{12}^{\mu}+\frac{i}{2}\theta_{1}^{a\alpha}(\gamma^{\mu})_{\alpha}^{\,\,\beta}\theta^a_{2\beta},\\
(\tilde{X}_{12})_{\alpha}^{\,\,\beta}=(X_{12})_{\alpha}^{\,\,\beta}+i\theta^a_{1\alpha}\theta_{2}^{a\beta}
+\frac{i}{2}(\theta^a_{1}\theta^a_{2})\delta_{\alpha}^{~\beta}.
\end{gather}

Super-Poincare invariant structure can thus be easily constructed out of these. However, as may easily be checked, they do not transform covariantly under superconformal transformations generated by $S^a_\alpha $. The alternative definition $S^a_\alpha = I Q^a_\alpha I$ will be useful for us.

In \cite{Park1999cw}, the following structures were defined and determined to transform homogeneously under super-inversion
\begin{equation}
(X_{ij+})_{\alpha}^{\,\,\beta}=(X_{i+})_{\alpha}^{\,\,\beta}-(X_{j-})_{\alpha}^{\,\,\beta}+i\theta_{i\alpha}^{a}\theta_{j}^{\,\beta a},
\end{equation}
\begin{equation}
(X_{ij-})_{\alpha}^{\,\,\beta}=(X_{i-})_{\alpha}^{\,\,\beta}-(X_{j+})_{\alpha}^{\,\,\beta}-i\theta_{j\alpha}^{a}\theta_{i}^{\,\beta a},
\end{equation}
with the transformation
\begin{gather}
I \left(X_{ij-} \right)_{\alpha}^{\,\,\beta} = -(X^{-1}_{j+})_\alpha^{~\gamma} (X_{ij-})_\gamma^{~\delta}(X^{-1}_{i-})_\delta^{~\beta},\\
I \left(X_{ij-} \right)_{\alpha}^{\,\,\beta} = -(X^{-1}_{i+})_\alpha^{~\gamma} (X_{ij+})_\gamma^{~\delta}(X^{-1}_{j-})_\delta^{~\beta}.  
\end{gather}

These can be written in the alternative form
\begin{equation}
X_{ij\pm}=\tilde{X}_{ij}\pm \frac{i}4\theta_{ij}^{2}\mathds{1},
\end{equation} 
which makes it clear that a particular combination of the above two objects annihilated by the supersymmetry generators has good super-inversion properties.

We also have the relations
\begin{gather} X_{i+}X_{i-}=\bar x_i^2\mathds{1},\\
X_{ij+}X_{ij-}=\bar{x}_{ij}^2 \mathds{1},
\end{gather}
where we have defined  
\begin{equation}
\bar x_i^2=x_i^2+\frac1{16}(\theta^a_i\theta^a_i)^2,\quad \bar{x}_{ij}^2= \tilde{x}_{ij}^{2}+\frac{1}{16}(\theta_{ij}^{a}\theta_{ij}^{a})^{2},
\end{equation}
which will be used repeatedly below. We have the following homogeneous transformation law under super-inversion:
\be
\bar x_{ij}^2 \xrightarrow{\phantom{~~}i_{s}\phantom{~~}} \frac{ \bar x_{ij}^2}{\bar x_i^2 \bar x_j^2}
\ee

With three superspace points, one can also construct  
\begin{equation}\label{Chi}
\mathfrak{X}_{1+}=X_{12-}^{-1}X_{23+}X_{31-}^{-1},
\end{equation}
with again a homogeneous transformation under super-inversion,
\begin{equation}
\mathfrak{X}_{1+}=X_{12-}^{-1}X_{23+}X_{31-}^{-1}\xrightarrow{\phantom{~~}i_{s}\phantom{~~}} -X_{1-}\mathfrak{X}_{1+}X_{1+},\end{equation}
and similarly for $\mathfrak{X}_{2+},\mathfrak{X}_{3+}$.

Superspace also admits purely fermionic 3-point superconformal covariant structures \cite{Park1999cw} defined as
\begin{equation}
\Theta_{1\alpha}^{a}=\left((X_{21+}^{-1}\theta_{21}^a)_{\alpha}-(X_{31+}^{-1}\theta_{31}^a)_{\alpha}\right),
\end{equation}
and transforming under super-inversion as
\begin{equation}
\Theta_{i\alpha}^{a}\xrightarrow{\phantom{~~}i_{s}\phantom{~~}}-(X_{i-})_{\alpha}^{\,\,\beta}\Theta_{i\beta}^{b}V_{ib}^{\,\,\,a}\,\,\,\,\,\,\,\,\Theta_{i}^{a\alpha}\xrightarrow{\phantom{~~}i_{s}\phantom{~~}} V^{Ta}_{i\,\,b}\,\Theta_{i}^{\beta\,b}(X_{i+})_{\beta}^{\,\,\alpha},
\end{equation}
with $\Theta_{2},\,\Theta_{3}$ defined similarly. Here $V_j, \,\,j=1,2,3$ is a matrix acting on the R-symmetry indices and has the form \cite{Park1999cw}
\begin{equation}
V_{jb}^{\,\,\,\,a} \equiv \delta_{b}^{a}+ i \theta_{jb}X_{j+}^{-1}\theta^{a}_{j}.
\end{equation}
It satisfies $V_j^T V_j=I$. As we will see in the next section, the existence of $V$ makes it non-trivial to construct super-inversion invariants which are also R-symmetry singlets.\footnote{Note that $V$ is trivial for $\cN=1$ as there is no R-symmetry in this case.}

$X_{ij\pm}, \Theta_{i\alpha}^{a}$ are the primary building blocks, together with the polarisation spinors that we discuss next,  out of which 3-point invariants for spinning correlators will be constructed. 

\subsection{Polarisation spinors}
We will adopt the formalism of \cite{Giombi:2011rz} to encode the tensor structure of the correlators and use polarisation spinors ($\lambda_{\alpha}$) as trackers of the spin of the operators. Our $\lambda$'s are 2-component, real, bosonic objects that transform in the spinor representation of the 3d Lorentz group. We define
 \begin{equation} 
 O_{\alpha_{1}\alpha_{2}.....\alpha_{2s}}\equiv (\gamma^{\mu_{1}})_{\alpha_{1}\alpha_{2}}(\gamma^{\mu_{2}})_{\alpha_{3}\alpha_{4}}....(\gamma^{\mu_{s}})_{\alpha_{2s-1}\alpha_{2s}} O_{\mu_{1}\mu_{2}.....\mu_{s}},
\end{equation} 
and 
 \begin{equation} 
O_s\equiv O_{\alpha_{1}\alpha_{2}.....\alpha_{2s}}\lambda^{\alpha_1}\lambda^{\alpha_2}.....\lambda^{\alpha_{2s}},
\end{equation} 
as a scalar object (with the spin encoded in the $\lambda$'s) representing a spin $s$ symmetric superconformal primary operator. We will be interested in 3-point functions of the form 
\be
\langle O_{s_{1}}(x_1, \theta_1, \lambda_1) O_{s_{2}}(x_2, \theta_2, \lambda_2) O_{s_{3}}(x_3, \theta_3, \lambda_3)\rangle \nonumber
\ee
with a homogeneity in $\lambda$ of $\lambda_1^{2s_1}\lambda_2^{2s_2}\lambda_3^{2s_3}$. This correlator will thus be a function of $(x_i,\theta_i, \lambda_i),\,\, (i=1,2,3)$, which are co-ordinates in (extended) superspace.\footnote{Note that $O_s(x,\theta)$ represents a local operator {\it superfield}, i.e. it is an operator super-multiplet with both bosonic and fermionic operators as components.}

Their transformation under super-inversions is the same as that of $\theta$'s :
\begin{equation} 
\lambda_{\alpha}\rightarrow(X_{+}^{-1}\lambda)_{\alpha}\,\,\,\,\,,\,\,\,\,\,\,
\lambda^{\beta}\rightarrow-(\lambda X_{-}^{-1})^{\beta}.
\end{equation}

In higher dimensions, where we have mixed symmetry tensor operators, one has to include fermionic polarisation spinors as well.

% move to appendix
\subsection{Supercovariant derivative and chiral co-ordinates}
In studying conservation constraints in Section \ref{CC} we will make much use of the supercovariant derivative $D^{a}_{\alpha}$ which as a differential operator in superspace is defined as
\begin{align}
D^{a}_{\alpha}\equiv\frac{\partial}{\partial \theta^{a\alpha}}+\frac i2\theta^{a\beta}\partial_{\beta\alpha},
\end{align}
where $\partial_{\beta\alpha}=(\gamma^{\mu})_{\beta\alpha}\partial_{\mu}$.

As is well-known, the operator $D^{a}_{\alpha}$ anti-commutes with all the supersymmetry generators
\begin{align}
\{D^{a}_{\alpha},Q^{b}_{\beta}\}=0.
\end{align}
Also,
\begin{align}
\{D^{a}_{\alpha},D^{b}_{\beta}\}=-P_{\alpha\beta}\delta^{ab}.
\end{align}
where $P_{\alpha\beta}=P^{\mu}(\gamma_{\mu})_{\alpha\beta}$.

\subsubsection*{Chiral basis for ${\cal N}=2$}

The $R$-symmetry group for ${\cal N}=2$ is $SO(2)$, which is isomorphic to $U(1)$. Thus, instead of working with real coordinates $\theta^{a\alpha}=(\theta^{1\alpha},\theta^{2\alpha})$, one can introduce the following chiral basis,
\begin{align}
\theta^{\alpha}\equiv\frac1{\sqrt{2}}(\theta^{1\alpha}+i\theta^{2\alpha})\ ,\qquad
\bar\theta^{\alpha}\equiv\frac1{\sqrt{2}}(\theta^{1\alpha}-i\theta^{2\alpha}).
\end{align}

The corresponding supercovariant derivatives can then be written as 
\begin{align}
D_{\alpha}=\frac{1}{\sqrt{2}}(D^{1}_{\alpha}-iD^{2}_{\alpha})\ ,\qquad
\bar D_{\alpha}=-\frac{1}{\sqrt{2}}(D^{1}_{\alpha}+iD^{2}_{\alpha}).
\end{align}

Expanding, one gets
\begin{align}
D_{\alpha}=\frac{\partial}{\partial \theta^{\alpha}}+\frac i2\bar\theta^{\beta}\partial_{\beta\alpha}\ ,\qquad 
\bar D_{\alpha}=\frac{\partial}{\partial \bar \theta^{\alpha}}+\frac i2\theta^{\beta}\partial_{\beta\alpha}.
\end{align}

Also, the (anti-)chiral supercovariant derivatives satisfy the algebra
\begin{align}
\{D_{\alpha},\bar D_{\beta}\}=-P_{\alpha\beta} ,\qquad \{D_{\alpha},D_{\beta}\}=\{\bar D_{\alpha},\bar D_{\beta}\}=0.
\end{align}

\section{The parity-even invariants and relations}\label{PE}

We will now determine 3-point invariant structures which will be used for writing down expressions for spinning correlators in $\cN=2$ 3d SCFTs. We will find new invariant structures, both parity-even and parity-odd. These new fermionic structures do not, of course, exist in the non-supersymmetric case \cite{Giombi:2011rz}, but they are also different from the $\cN=1$ case considered in \cite{Nizami:2013tpa}, as we will discover \footnote{One may wonder how there are additional invariants for 3-point function structures given that $\cN=2$ susy is more constraining than $\cN=1$. Firstly, with more grassmanian variables there can be a higher non-vanishing degree of $\Theta$ in an invariant. Specifically, for ${\cal N}=1$, two factors of $\Theta$ is the highest possible in a non-vanishing term, while for ${\cal N}=2$, invariants with $\mathcal{O} (\Theta^{4})$ (like $R'$ defined below, which do not have a counterpart in $\mathcal{N}=1$) are also possible. Also, note that since the size of the supermultiplet increases with $\mathcal{N}$, the superspace 3-point function will, with increasing $\mathcal{N}$, contain more and more `elementary' 3-point functions in its component expansion. Since more elementary correlators are contained within a single superspace correlator for larger $\mathcal{N}$, it is expected that a larger number of invariants is required to encapsulate the structure.}.

In this section, we will construct all the parity even 3-point superconformal invariants using the covariant structures of the previous section, and also the polarisation spinors encoding spin information. The bosonic\footnote{By bosonic we mean those invariants which are non-vanishing when all the grassmanian coordinates are put to zero. Likewise, fermionic invariants are defined as those that vanish identically when the grassmanian coordinates are set to zero, and thus exist only in supersymmetric theories. Note that one could have called ``fermionic"  the invariants which are used to build up the 3-point function structures of free/critical fermion CFTs (for example) with no susy - however this is not our usage. For us,  fermionic invariants exist only in superspace and vanish in ordinary (non-supersymmetric) CFTs. We thank the referee for a comment regarding this matter.} invariants are direct generalisations of the non-supersymmetric ($\mathcal{N}=0$) ones in \cite{Giombi:2011rz}
\begin{equation}\label{Pdef}
P_{1}\equiv \lambda_{2}X_{23-}^{-1}\lambda_{3}\,,\,\,\,\,\, P_{2}\equiv \lambda_{3}X_{31-}^{-1}\lambda_{1}\,,\,\,\,\,\, P_{3}\equiv \lambda_{1}X_{12-}^{-1}\lambda_{2}\end{equation}
\begin{equation}
Q_{1}\equiv \lambda_{1}\mathfrak{X}_{1+}\lambda_{1}\,,\,\,\,\,\, Q_{2}\equiv \lambda_{2}\mathfrak{X}_{2+}\lambda_{2}\,\,\,\,\,\, Q_{3}\equiv \lambda_{3}\mathfrak{X}_{3+}\lambda_{3}.\end{equation}
where $\mathfrak{X}_{i+}$ is defined in Eq. \eqref{Chi}. Using the transformations of the covariant structures under superinversion listed in Section \ref{Scov}, it is straightforward to check that these are parity-even superconformal invariants.

The fermionic parity-even invariants in the $\mathcal{N}=1$ case (with no R-symmetry) are \cite{Nizami:2013tpa} $R_i=\lambda_i \T_i, \,\, i=1,2,3$. For the $\cN =2$ case these objects will be charged under R-symmetry: $R_i^a=\lambda_i \T_i^a$. However, in this case R-symmetry singlets like $R_i^2$ are vanishing, whereas those like $R_1\cdot R_2 = \delta_{ab} R_1^a R_2^b$ are not super-inversion invariant because of the non-trivial $V$ factors in the transformation of $\T$'s. Thus we can not straightforwardly build on the $\cN=1$ fermionic invariants to get the $\cN=2$ ones and must undertake an {\it ab-initio} construction. 

\subsection{Fermionic counterparts of the $P, Q, S$ invariants}
As a first step towards constructing the fermionic invariants, we will first define a number of covariant fermionic counterparts to the $P, Q, S$ invariants above by replacing one or more $\lambda$'s by $\Theta$. The requisite definitions, together with the transformation under super-inversion ($i_s$), are
\begin{align}
\pi_{ij}^{a}&=\lambda_{i}X_{ij+}\Theta_{j}^{a}\xrightarrow{\phantom{~~}i_{s}\phantom{~~}} -\dfrac{1}{\bar{x}_{i}^{2}}\pi_{ij}^{b} (V_{j})_{b}^{\ a}, \\
\vspace{1em}
\Pi_{ij}^{ab}&=\Theta_{i}^{a}X_{ij+}\Theta^{b}_{j}\xrightarrow{\phantom{~~}i_{s}\phantom{~~}} (V^{T}_{i})^{a}_{\ c}\Pi_{ij}^{cd}(V_{j})_{d}^{\ b},\\
\vspace{1em}
\omega^{a}_{i}&=\lambda_{i}\mathfrak{X}_{i+}\Theta_{i}^{a}\xrightarrow{\phantom{~~}i_{s}\phantom{~~}} -\bar{x}_{i}^{2}\omega_{i}^{b}(V_{i})_{b}^{\ a},\\
\vspace{1em}
\Omega_{i}^{ab}&=\Theta_{i}^{a}\mathfrak{X}_{i+}\Theta_{i}^{b}\xrightarrow{\phantom{~~}i_{s}\phantom{~~}} \bar{x}_{i}^{4}(V^{T}_{i})^{a}_{\ c}\Omega_{i}^{cd}(V_{i})_{d}^{\ b},\\
\vspace{1em}
\sigma_{13}^{a}&=\dfrac{\lambda_{1}X_{12+}X_{23+}\Theta_{3}^{a}}{\sqrt{\bar x_{12}^{2}\bar x_{23}^{2}\bar x_{31}^{2}}}\xrightarrow{\phantom{~~}i_{s}\phantom{~~}} \bar{x}_{3}^{2}\sigma_{13}^{b}(V_{3})_{b}^{\ a} ,\\
\vspace{1em}
\Sigma_{13}^{ab}&=\dfrac{\Theta^{a}_{1}X_{12+}X_{23+}\Theta^{b}_{3}}{\sqrt{\bar x_{12}^{2}\bar x_{23}^{2}\bar x_{31}^{2}}}\xrightarrow{\phantom{~~}i_{s}\phantom{~~}} - \bar{x}_{1}^{2}\bar{x}_{3}^{2} (V^{T}_{1})^{a}_{\ c}\Sigma^{cd}_{13}(V_{3})_{d}^{\ b}.
\end{align}
These objects are charged under R-symmetry {\it and} are superconformal {\it co}variant - they transform homogeneously under super-inversion.\footnote{Cyclic permutations of the last two structures above can be written down straightforwardly.}

We can now build the required  invariants by suitable contraction of the various covariant fermionic structures constructed above. However, not all R-symmetry singlets constructed out of the above covariant structures will be invariant under super-inversion. This subtlety is due to the existence of the $V$ matrix in R-symmetry space. Let us take a simple example to explain this feature. A natural choice for a fermionic parity-even invariant\footnote{Note that for $\cN=1$, both $R_1, R_2$ are separately invariant and thus so is $R_1 \cdot R_2$.} would be $R_1\cdot R_2 \equiv R_1^a R_2^b \delta_{ab}$. However, under superinversion  we get the $V$ matrix at points 1 and 2 and thus superconformal invariance is not maintained. On the other hand, consider the R-symmetry singlet $ \frac{1}{\bar{x}_{12}^{2}}\pi_{13}^{a}\pi_{23}^{a}$ which, like $R_1\cdot R_2$, is also of homogeneity $\lambda_1 \lambda_2$. In this case, one can check that super-inversion invariance is maintained.

We thus have the following set of parity-even fermionic invariants
\begin{equation}\label{Rbdef} 
\Rb_{1}\equiv \frac{1}{\bar{x}_{23}^{2}}\pi_{21}^{a}\pi_{31}^{a},\,\,\,\,\Rb_{2}\equiv \frac{1}{\bar{x}_{31}^{2}}\pi_{32}^{a}\pi_{12}^{a},\,\,\,\,\Rb_{3}\equiv \frac{1}{\bar{x}_{12}^{2}}\pi_{13}^{a}\pi_{23}^{a}. \end{equation}

We also have a fermionic parity-even invariant without any $\lambda$'s (such an object does not exist in the $\cN=1$ case)
\begin{equation}\label{Rpdef}
R' \equiv \Pi_{ij}^{ab}\Pi_{ij}^{ab}
\end{equation}
Here we have a sum over $a,b$ but no sum over $i,j$. It is clearly neutral under R-symmetry and it can be checked that it is super-inversion invariant as well.\footnote{Note that $i \neq j$. It turns out that the three different pairs of $(i,j)$ are all equivalent and thus there is only one $R'$ invariant which needs to be defined here.} The choice of $\Rb_i$ and $R'$ may seem arbitrary. For example,  we can also build the following object by contracting the above covariant structures: $\dfrac{\bar x_{23}^{2}\bar x_{31}^{2}}{\bar x_{12}^{2}}\sigma_{13}^{a}\sigma_{23}^{a}$ . This R-symmetry neutral object (or its cyclic permutations) can be checked to be parity-even and super-inversion invariant, and one may wonder if it defines a new parity-even fermionic invariant. Likewise, one could ask the same question for the parity-even, R-symmetry singlet, superconformally invariant structure $\frac{\bar x_{12}^{2}\bar x_{31}^{2}}{\bar x_{23}^{2}}\Omega^{aa}_{1}$. However, one can show that
\be
\frac{\bar x_{12}^{2}\bar x_{31}^{2}}{\bar x_{23}^{2}}\Omega^{aa}_{1}=\frac{i}{2}R',\, \quad \dfrac{\bar x_{23}^{2}\bar x_{31}^{2}}{\bar x_{12}^{2}}\sigma_{13}^{a}\sigma_{23}^{a}=\bar R_{3}-\dfrac i2 R' P_{3},
\ee
so that both these cases can be expressed in terms of the existing set of parity-even invariants. There are many more examples like these. A systematic analysis (see Appendix \ref{rel}) shows that we can work with the set of invariants $P_i, Q_i, \Rb_i$ and $R'$, and all the parity-even structures that are constructed from the above covariant building blocks can be expressed in terms of these. 

\subsection{Relations for even invariants}
To summarise, the parity-even invariants are $P_i, Q_i, \Rb_i$ and $R'$. With the action on three superspace points of the $\cN=2$ superconformal group in 3d, with 19 generators (11 bosonic and 8 fermionic), we expect to be able to construct $ 9 \times 3-19=8 $  such independent invariants. Thus there should be two relations between the 10 constructed invariants. Indeed we find,\footnote{These extend the corresponding relations in \cite{Giombi:2011rz, Nizami:2013tpa}.}
\begin{align}\label{GPE}
\sum_{i=1}^{3}P_{i}^{2}Q_{i}-2P_{1}P_{2}P_{3}-Q_{1}Q_{2}Q_{3}+\frac i2\sum_{i=1}^{3}\bar R_{i}P_{i}Q_{i}+\frac14 R' P_{1}P_{2}P_{3}=0,
\end{align}
at ${\cal O}(\lambda_{1}^{2}\lambda_{2}^2\lambda_{3}^{2})$, and the (cyclically related) triplet of relations at ${\cal O}(\lambda_{i}^{2}\lambda_{j}\lambda_{k})$:
\begin{align}\label{FPE}
P_{1}\bar R_{2}+P_{2}\bar R_{1}+Q_{3}\bar R_{3}-\frac i2 R'P_{1}P_{2}&=0, \nonumber\\
P_{2}\bar R_{3}+P_{3}\bar R_{2}+Q_{1}\bar R_{1}-\frac i2 R'P_{2}P_{3}&=0,\\
P_{3}\bar R_{1}+P_{1}\bar R_{3}+Q_{2}\bar R_{2}-\frac i2 R'P_{3}P_{1}&=0 \nonumber
\end{align}
Together, these make the number of independent parity-even invariants to be eight, as required.

Other relations between the parity-even invariants, which are higher order in the $\lambda$'s, can be generated from the above relations. Some of these are listed at the end of Appendix \ref{rel}. 

\section{The parity-odd invariants and relations} \label{PO}
In this section we turn to the construction of 3-point function invariants which are odd under parity. These pick up a minus sign under a super-inversion transformation. As previously, there are some of these which are direct extensions of the $\cN=0$ case \cite{Giombi:2011rz} :
\begin{equation}
S_{1}=\frac{\lambda_{3}X_{31+}X_{12+}\lambda_{2}}{\sqrt{\bar x^{2}_{12}\bar x^{2}_{23}\bar x^{2}_{31}}}\,,\,\,\,\,\, S_{2}=\frac{\lambda_{1}X_{12+}X_{23+}\lambda_{3}}{\sqrt{\bar x^{2}_{12}\bar x^{2}_{23}\bar x^{2}_{31}}}\,,\,\,\,\,\, S_{3}=\frac{\lambda_{2}X_{23+}X_{31+}\lambda_{1}}{\sqrt{\bar x^{2}_{12}\bar x^{2}_{23}\bar x^{2}_{31}}}.\end{equation}
Given the transformation laws of the building blocks in Section \ref{Scov}, it is easy to check that these transform as $S_i \rightarrow -S_i$ under super-inversion and hence are parity-odd.

Also, there now exists a unique parity-odd fermionic invariant:
\begin{equation}
T'=\sqrt{\dfrac{\bar x_{12}^{2}\bar x_{31}^{2}}{\bar x_{23}^{2}}}\displaystyle\Theta^{a}_{1\alpha}\Theta^{a\alpha}_{1}	\end{equation}

Under super-inversion: $T'\rightarrow -T'$.\footnote{It can be checked that cyclic permutations of $1,2,3$ give the same invariant.}

Like in the parity-even case, there are other parity-odd invariant structures that can be built from the covariant building blocks of Section \ref{Scov} but they can all be expressed in terms of $S_i$ and $T'$(see Appendix \ref{rel}). % \footnote{To find the fermionic parity-odd invariants..} 

\subsection{Relations for odd invariants}
Since the product of two odd invariants is an even one, we must be able to express it in terms of the minimal set of even invariants. For example, we have
\begin{align}
T'S_{i}=-2\bar R_{i}.
\end{align}

We further have relations containing products of $S_{i}$'s which is expected to be an even invariant. Indeed, we find
\begin{align}
S_{1}^{2}=P_{1}^{2}-Q_{2}Q_{3}+i\bar R_{1}P_{1},\quad S_{1}S_{2}=-P_{1}P_{2}+P_{3}Q_{3},
\end{align}
and cyclic permutations.

Similarly for $T'$, we get
\begin{align}
T'^{2}=2R'.
\end{align}

We also have various linear relations between products of even and odd invariants.

\begin{itemize}
\item ${\cal O}(\lambda_1\lambda_2)$:
    \begin{gather}
    \bar R_{i}T'=-S_{i}R' 
    \end{gather}

    \item ${\cal O}(\lambda_1^2\lambda_2^2)$:
    \begin{gather}\label{PTrel}
    P_3^2T'-Q_1Q_2T'+2S_3\bar R_3-iP_3S_3R'=0,
    \end{gather}
    and cyclic permutations.
    
    \item {${\cal O}(\lambda_{1}^{2}\lambda_{2}\lambda_{3})$}:
	\begin{gather}
	Q_{1}S_{1}+P_{2}S_{3}+P_{3}S_{2}+\frac i2 T'P_{2}P_{3}=0,\\
	S_{2}\bar R_{3}-S_{3}\bar R_{2}=0\label{SRrel},\\
	S_{2}\bar R_{3}+S_{3}\bar R_{2}+T'(P_1Q_1-P_2P_3)=0,
	\end{gather}
and cyclic permutations.

    \item {${\cal O}(\lambda_{1}^{2}\lambda_{2}^{2}\lambda_{3}^{2})$}:
	\begin{gather}
	P_{1}Q_{1}S_{1}+P_{2}Q_{2}S_{2}-P_{3}Q_{3}S_{3}+2P_{1}P_{2}S_{3}+\frac i2 T'P_{1}P_{2}P_{3}=0,\\
	\bar R_{1}(
	S_{2}P_{3}+\frac12 Q_{1}S_{1}
	)+\bar R_{2}(
	S_{3}P_{1}+\frac12 Q_{2}S_{2}
	)+\bar R_{3}(
	S_{1}P_{2}+\frac12 Q_{3}S_{3}
	)+\frac i4 R'S_{1}S_{2}S_{3}=0,
	\end{gather}
and cyclic permutations.

    \item {${\cal O}(\lambda_{1}^{3}\lambda_{2}^{3}\lambda_{3}^{2})$}:
	\begin{gather}
	(P_{1}^{2}Q_{1}-P_{2}^{2}Q_{2})P_{3}S_{3}+(P_{3}^{2}-Q_{1}Q_{2}+iP_{3}\bar R_{3})(P_{1}Q_{1}S_{1}-P_{2}Q_{2}S_{2})=0,
	\end{gather}
and cyclic permutations. 

\end{itemize}
Like in the parity-even case, these relations between the various parity-odd structures have to be taken into account when enumerating all the independent structures for the 3-point correlators in terms of our minimal set of invariants.

\section{Three-point function examples}\label{3pt}

The purpose of constructing the parity even and odd invariants is to make simpler the enumeration of tensor structures for particular 3-point correlators, a task we turn to in this section. 
Before turning to specific examples of 3-point functions of spinning operators, let us make some general observations.

The general 3-point function of spinning operators has the following form in terms of the invariants constructed in the previous sections:
\begin{equation}
\langle O_{s_{1}}(x_1, \theta_1,\lambda_1)O_{s_{2}}(x_2, \theta_2, \lambda_2)O_{s_{3}}(x_3, \theta_3, \lambda_3)\rangle =\frac{1}{\bar{x}_{12}^{m_{123}}\bar{x}_{23}^{m_{231}}\bar{x}_{31}^{m_{312}}}\sum_{n}\mathcal{G}_{n}(P_{i}, Q_{i},\bar R_{i},R', S_{i},T'),
\end{equation}
with $m_{ijk}\equiv (\Delta_{i}-s_{i})+(\Delta_{j}-s_{j})-(\Delta_{k}-s_{k})$,\footnote{$\Delta_i$ and $s_i$ denote, respectively, the scaling dimension and spin of the operator $O_s$} the sum being over all the {\it linearly independent} invariant structures $\mathcal{G}_{n}$,
each with homogeneity $\lambda_{1}^{2s_{1}}\lambda_{2}^{2s_{2}}\lambda_{3}^{2s_{3}}$.
Now the 3-point function has to be linear in the parity-odd invariants and in  $R'$ or $\bar R_i$'s. We thus have the following general structure for $\mathcal{G}_{n}$:
\begin{align*}
\mathcal{G}_{n}=G_{n}^{(0)}(P_{i},Q_{i})+a_{n}^{(1)}G_{n}^{(0)}(P_{i},Q_{i})R'+a_{n}^{(2)}G_{n}^{(1)}(P_{i},Q_{i})\bar R_{j} +b_{n}^{(1)}G_{n}^{(0)}(P_{i},Q_{i})T'\\
+b_{n}^{(2)}G_{n}^{(1)}(P_{i},Q_{i})S_{k}+b_{n}^{(3)}G_{n}^{(1)}(P_{i},Q_{i})R'S_{k}+b_{n}^{(4)}G_{n}^{(2)}(P_{i},Q_{i})\bar R_{j}S_{k},
\end{align*}
where $G_{n}^{(a)}(P_{i},Q_{i})$ is a monomial in $P$'s and $Q$'s with every term in the r.h.s above having homogeneity $\lambda_{1}^{2s_{1}}\lambda_{2}^{2s_{2}}\lambda_{3}^{2s_{3}}$.
 
 The coefficients of the monomials can be further constrained when one or more operators are conserved currents, as we will see in a later section. Besides this, constraints also arise when two or more operators are identical, giving rise to permutation invariance in the correlator. Our invariants have the following transformation under a swap of superspace points 2 and 3:
 \begin{gather*}
 A_1 \rightarrow -A_1\qquad A_2 \rightarrow -A_3,\qquad A_3 \rightarrow -A_2, \\
 R' \rightarrow R',\qquad T' \rightarrow T',
 \end{gather*}
 where $A_i$ stands for any of $P_i, Q_i,\bar R_i, S_i$. This will be useful when enumerating correlator structures below.
 
\subsection{Examples}
\subsection*{ $\la O_s O_0 O_0 \ra$}
This correlator of one spin-$s$ and two scalar operators has homogeneity $\lambda_{1}^{2s}$ and thus the possible structures are
\begin{align}
 Q_1^s, \,\,\,Q_1^s R',\,\,\,Q_1^s T'.
\end{align}
 In this case all the 3 structures are independent, so the general form of this correlator is
\begin{align}
\langle O_s O_0 O_0\ra=\frac{1}{\bar{x}_{12}^{m_{123}}\bar{x}_{23}^{m_{231}}\bar{x}_{31}^{m_{312}}} \Big(a_1 Q_1^s+a_2 Q_1^s R'+b_1Q_1^s T'\Big),\quad s=0,2,4,\hdots.
\end{align}
 
Further we note that when $s$ is an odd integer, each of  the 3 structures above pick up a minus sign under $2\leftrightarrow3$ whereas the correlator is symmetric under this swap. Thus, we can conclude that $\la O_s O_0 O_0 \ra$ vanishes for $s$ odd.

 \subsection*{$\la O_1 O_1 O_0 \ra$}
The possible structures for the correlator of two identical spin one operators and one scalar operator (homogeneity $\lambda_1^2\lambda_2^2$) are:
\begin{align}
\begin{aligned}
&P_3^2, \,\,\,\,Q_1 Q_2,\,\,\,P_3^2 R',\,\,\,Q_1 Q_2 R',\,\,\, P_3 \Rb_3\,\,\,\text{(parity even)} \\
 &Q_1 Q_2 T',\,\,\, P_3S_3,\,\,\, P_3 S_3 R',\,\,\, \Rb_3S_3\,\,\,\,\text{(parity odd)}
\end{aligned}
\end{align}
The structure $P_3^2 T'$ is also possible but can be eliminated using Eq. \eqref{PTrel}.

 \subsection*{$\langle
O_{s}O_{1}O_{0}
\rangle$}
The possible structures for this correlator are:
 \begin{align}
\begin{aligned}
&Q_{1}^{s}Q_{2},\  Q_{1}^{s}Q_{2}R', \ Q_{1}^{s-1}P_{3}^{2},\  Q_{1}^{s-1}P_{3}^{2}R',\  Q_{1}^{s-1}P_{3}\bar R_{3} \quad \text{(parity even)}\\[.5em]
&Q_{1}^{s}Q_{2}T',\ Q_{1}^{s-1}P_{3}S_{3},\ Q_{1}^{s-1}P_{3}S_{3}R',Q_{1}^{s-1}\bar R_{3}S_{3}\quad\text{(parity odd)}
\end{aligned}
\end{align}
We thus note that the possible structures for $ \la O_s O_1 O_0\ra $   are just $Q_1^{s-1}$ times the possible structures for $\la O_1 O_1 O_0\ra$.

\subsection*{$\langle
O_{s}O_{\frac12}O_{0}
\rangle$}
	
Only half-integral values of $s$ are allowed, since there is no invariant in the ${\cal N}=2$ theory with homogeneity ${\cal O}(\lambda)$.The allowed structures are:
\begin{align}
\begin{aligned}
&Q_{1}^{s-\frac12}P_{3},\ Q_{1}^{s-\frac12}P_{3}R',\ Q_{1}^{s-\frac12}\bar R_{3}\quad\text{(parity even)}\\
&Q_{1}^{s-\frac12}P_{3}T',Q_{1}^{s-\frac12}S_{3},\ Q_{1}^{s-\frac12}S_{3}R',\quad\text{(parity odd)}
\end{aligned}
\end{align}

\subsection*{$\langle
O_{2}O_{2}O_{0}
\rangle$}
The possible independent structures for this correlator are:
\begin{align}
\begin{aligned}
&Q_{1}^{2}Q_{2}^{2},\  Q_{1}^{2}Q_{2}^{2}R', \ Q_{1}Q_{2}P_{3}^{2},\  Q_{1}Q_{2}P_{3}^{2}R',\  Q_{1}Q_{2}P_{3}\bar R_{3},\  \\
&P_{3}^{4}, \ P_{3}^{4}R',\ P_{3}^{3}\bar R_{3}\quad\text{(parity even)}\\[.5em]
&Q_{1}^{2}Q_{2}^{2}T', \ Q_{1}Q_{2}P_{3}^{2}T',\ Q_{1}Q_{2}S_{3}P_{3},\ Q_{1}Q_{2}S_{3}P_{3}R',\ Q_{1}Q_{2}S_{3}\bar R_{3}\\
& P_{3}^{3}S_{3},\ P_{3}^{3}S_{3}R',\ P_{3}^{2}S_{3}\bar R_{3}\quad\text{(parity odd)}
%P_{3}^{4}T'
\end{aligned}
\end{align}
Again, some possible structures have been eliminated using the relations.

\subsection*{$\langle
O_{s}O_{2}O_{0}
\rangle$}
Once again, the possible structures for $ \la O_s O_2 O_0\ra $   are just $Q_1^{s-2}$ times the above possible structures for $\la O_2 O_2 O_0\ra$.

\subsection*{$\langle
O_{1}O_{1/2}O_{1/2}
\rangle$}
The only possible structures with the required symmetry under $2 \leftrightarrow3$ are parity odd
\be
P_2S_3-P_3S_2
\ee
The structure $\Rb_2S_3-\Rb_3S_2$ is also possible but vanishes by Eq. \eqref{SRrel}.

\subsection*{$\langle
O_{2}O_{1/2}O_{1/2}
\rangle$}
The independent structures are
\begin{align}
\begin{aligned}
&Q_{1}P_{2}P_{3},\ Q_{1}P_{2}P_{3}R',\ Q_{1}^{2}P_{1},\ Q_{1}P_{1}^{2}R',\ Q_{1}^{2}\bar R_{1}\qquad\text{(parity even)}\\[.5em]
&Q_{1}P_{2}P_{3}T',\ Q_{1}^{2}P_{1}T',\ Q_{1}^2 S_1,\ Q_{1}^2 S_1R' \qquad \text{(parity odd)}
\end{aligned}
\end{align}
Under $2 \leftrightarrow 3$ this correlator picks up a minus sign because of the anti-commuting fermionic operators. Other possible structures include $Q_1(P_2\Rb_3+P_3\Rb_2),\,\,Q_1(P_2S_3+P_3S_2),\,\, Q_1(P_2S_3+P_3S_2)R'$ but can be expressed in terms of the above structures by using the relations in Sections \ref{PE} and \ref{PO}.
\subsection*{$\langle
O_{s}O_{1/2}O_{1/2}
\rangle$}
The possible structures for this correlator, for $s$ even, are $Q_1^{s-2}$ times the allowed structures for $\la O_{2}O_{1/2}O_{1/2} \ra$. For $s$ odd, they are $Q_1^{s-1}(P_2S_3-P_3S_2)$. 
%times the allowed structures for $\la O_{1}O_{1/2}O_{1/2} \ra$.

\subsection*{$\langle
O_{2}O_{1}O_{1}
\rangle$}
The independent structures are
\begin{align}
\begin{aligned}
&Q_{1}^{2}Q_{2}Q_{3},\ Q_{1}^{2}Q_{2}Q_{3}R',\ Q_{1}^{2}P_{1}^{2},\ Q_{1}^{2}P_{1}^{2}R',\ Q_{1}P_{1}P_{2}P_{3},\ Q_{1}P_{1}P_{2}P_{3}R',\ P_{2}^{2}P_{3}^{2},\\
& P_{2}^{2}P_{3}^{2}R',\ Q_{1}^{2}P_{1}\bar R_{1},\ Q_{1}P_{2}P_{3}\bar R_{1}\qquad\text{(parity even)}\\[.5em]
&Q_{1}^{2}Q_{2}Q_{3}T',\ Q_{1}P_{1}P_{2}P_{3}T',\ Q_{1}^{2}P_{1}S_{1},\ Q_{1}^{2}P_{1}S_{1}R',\\
&Q_{1}P_{2}P_{3}S_{1},\ Q_{1}P_{2}P_{3}S_{1}R',\ Q_{1}^{2}S_{1}\bar R_{1},\ P_{2}^{2}S_{3}\bar R_{3}+P_{3}^{2}S_{2}\bar R_{2}\qquad \text{(parity odd)}
%\ Q_{1}^{2}P_{1}^{2}T', P_{2}^{2}P_{3}^{2}T'
\end{aligned}
\end{align}
These have the requisite symmetry under a $2 \leftrightarrow 3$ swap. Other structures, which are possible on grounds of homogeneity and symmetry can be expressed in terms of the above list by using the relations in Sections \ref{PE} and \ref{PO}.

\subsection*{$\langle
O_{s}O_{1}O_{1}
\rangle$} The possible independent structures for this correlator are simply $Q_1^{s-2}$ times the structures for $\langle O_{2}O_{1}O_{1} \rangle$ above, for $s=2,3,...$.

\subsection*{$\langle
O_{2}O_{2}O_{2}
\rangle$}
This correlator is fully symmetric under $1\leftrightarrow2 \leftrightarrow 3$, and the independent structures are
\begin{align}
\begin{aligned}
&Q_{1}^{2}Q_{2}^{2}Q_{3}^{2},\ Q_{1}^{2}Q_{2}^{2}Q_{3}^{2}R',\ P_{1}^{2}P_{2}^{2}P_{3}^{2},\ P_{1}^{2}P_{2}^{2}P_{3}^{2}R',\ Q_{1}Q_{2}Q_{3}P_{1}P_{2}P_{3},\ Q_{1}Q_{2}Q_{3}P_{1}P_{2}P_{3}R'  \\[.5em]
&\sum_{i}P_{i}^{4}Q_{i}^{2},\ R'\sum_{i}P_{i}^{4}Q_{i}^{2},\ P_{1}P_{2}P_{3}\sum_{\text{cyc}}P_{1}P_{2}\bar R_{3},\ Q_{1}Q_{2}Q_{3}\sum_{\text{cyc}}P_{1}P_{2}\bar R_{3}\quad \text{(parity even)}\\[1em]
&Q_{1}^{2}Q_{2}^{2}Q_{3}^{2}T',\ P_{1}^{2}P_{2}^{2}P_{3}^{2}T',\ Q_{1}Q_{2}Q_{3}P_{1}P_{2}P_{3}T',\ T'\sum_{i}P_{i}^{4}Q_{i}^{2},\ \sum_{i}P_{i}^{3}Q_{i}^{2}S_{i},\\
&R'\sum_{i}P_{i}^{3}Q_{i}^{2}S_{i},\ P_{1}P_{2}P_{3}\sum_{\text{cyc}}P_{1}P_{2}S_{3},\ R'P_{1}P_{2}P_{3}\sum_{\text{cyc}}P_{1}P_{2}S_{3} \qquad\text{(parity odd)}
\end{aligned}
\end{align}
Several other linearly dependent structures that are possible have been removed using various relations between the invariants.

\newpage
\section{Correlators of conserved currents}\label{CC}
Our analysis so far has been general and the only constraint on the operators/correlators has been  $\cN=2$ superconformal symmetry. In this section we specialise to the case when one or more operators $O_s$ in the 3-point function is a conserved current. We will denote a conserved current by $J_s$ which will obey the shortening condition  $D^a_{\alpha_1}J^{\alpha_1 \alpha_2 .....\alpha_{2s}}=0$ or equivalently,
\be \label{conseq}
D^{a\alpha} \frac{\partial}{\partial \lambda^{\alpha}} J_{s}=0
\ee
This is of interest because the stress-energy tensor is always an exactly conserved current in any SCFT, one can have (one or more) conserved spin one currents in the presence of global symmetries, and furthermore Chern-Simons-Matter theories have higher-spin currents for weakly broken higher-spin symmetries (when these symmetries are exact, the CFT is free \cite{Maldacena:2011jn}).The presence of conserved currents imposes further constraints on the structure correlators when the above conservation/shortening condition is imposed - some of the coefficients in the 3-point function are determined in terms of others as we will see below. For non-supersymmetric CFTs in various dimensions, \cite{Osborn:1993cr} had shown the existence of a fixed number (two in 3 dimensions) of (parity-even) tensor structures for the 3-point correlator of conserved currents. In \cite{Giombi:2011rz} it was shown that for 3d CFTs, conserved current correlators also admit a third structure which is parity-odd. For $\cN=1$ SCFTs, evidence was provided in \cite{Nizami:2013tpa} (see also \cite{Buchbinder:2015qsa, Buchbinder:2021qlb, Buchbinder:2021gwu}) that 3-point conserved current correlators have one parity-even and one parity-odd structure.

\subsection{Free field computation of conserved current correlators}
We will first focus on the special case of 3-point correlators in a free field $\cN=2$ SCFT. 
% Lagrangian
Such a theory possesses an infinite number of exactly conserved higher spin currents. The form of the currents in terms of the elementary (anti-) chiral superfield is \cite{Nizami:2013tpa} 
\begin{align}
J_{s}=\sum_{r=0}^{s}\left\{
(-1)^{r(2r+1)} \begin{pmatrix}
2s\\
2r
\end{pmatrix}\partial^{r}\bar\Phi\,\partial^{s-r}\Phi+(-1)^{(r+1)(2r+1)}\begin{pmatrix}
2s\\
2r+1
\end{pmatrix}\partial^{r}\bar D\bar \Phi\,\partial^{s-r-1}D\Phi
\right\}
\end{align}
where $\partial=i\lambda^{\alpha}(\gamma^{\mu})_{\alpha}\beta\lambda^{\beta}\partial_{\mu},\ D=\lambda^{\alpha}D_{\alpha}$, and $s=0,1,2\hdots$.

It is straightforward to directly compute 3-point correlators of such conserved currents by using Wick's theorem. The basic non-zero Wick contractions are between the chiral superfield and its conjugate, and given by
\begin{align}
\langle\bar\Phi_{1}\Phi_{2}\rangle=\frac{1}{y_{12}}
\end{align}
where $\Phi_{i}$ denotes the superfield at superspace point $x_{i}$ and $y_{12}$ is defined \cite{Park1999cw, Buchbinder:2015qsa} as
\begin{align}
y_{12}^{2}=\bar x_{12}^{2}v_{12},\quad v_{12}=\frac{1}{2}\left((V_{12})^{aa}+i\epsilon_{ab}(V_{12})^{ab}\right)
\end{align}
where $V_{12}^{ab}=\delta^{ab}+i\theta_{12}^{a\alpha}(X_{12+}^{-1})_{\alpha}^{\ \beta}\theta_{12\beta}^{b}$.

The computational procedure is explained in more detail through examples in Appendix \ref{FT}. Below we simply quote the final results.\footnote{The $f$ subscript denotes a free theory correlator. We have removed constant multiplicative overall factors.} As expected, we find that the result of the calculation can be expressed in terms of only the parity-even invariants,
\begin{align}
\langle
J_{0}J_{0}J_{0}
\rangle_f=
\frac{1}{\bar x_{12}\bar x_{23}\bar x_{31}}
\left(
1+\frac1{16}R'
\right)
\end{align}

\begin{align}
\langle
J_{2}J_{0}J_{0}
\rangle_f=
\frac{Q_{1}^{2}}{\bar x_{12}\bar x_{23}\bar x_{31}}
\left(1+\frac5{16}R'\right)
\end{align}

\begin{align}
\langle
J_{4}J_{0}J_{0}
\rangle_f=
\frac{Q_{1}^{4}}{\bar x_{12}\bar x_{23}\bar x_{31}}
\left(1+\frac9{16}R'\right)
\end{align}

\begin{align}
\langle J_{1}J_{1}J_{0}\rangle_f=\frac{1}{\bar x_{12}\bar x_{23}\bar x_{31}}\left[
Q_{1}Q_{2}\left(
1+\frac9{16}R'
\right)+P_{3}^{2}\left(
2+\frac{3}{8}R'
\right)+2iP_{3}\bar R_{3}
\right]
\end{align}

\subsection{General constraints of conservation for 3-point correlators}
We will now directly use the results of Section \ref{3pt} and impose conservation (Eq. \eqref{conseq}) on the correlator to obtain additional constraints. The general procedure is as follows. Since the 3-point correlator is a multinomial in the invariants,\footnote{Note that the exponents $m_{ijk}$ here are all unity for the case of conserved currents because the unitarity bound is saturated in this case.} 
\begin{equation} \label{J3form}
\langle J_{s_{1}}J_{s_{2}}J_{s_{3}}\rangle =\frac{1}{{\bar x}_{12}{\bar x}_{23}{\bar x}_{31}}\sum_{n}a_{n}\mathcal{G}_{n},\end{equation}
the imposition of the conservation constraints
\begin{equation}D_{i}^{a\alpha}\frac{\partial}{\partial\lambda _{i}^{\alpha}} \langle J_{s_{1}}J_{s_{2}}J_{s_{3}}\rangle=0,\quad i=1,2,3
\end{equation}
(no sum over $i$) will lead to linear relations between various coefficients of the multinomial. The computation gets rather messy. We set some superspace coordinates to special values, and worked through a number of examples on Mathematica. The results are tabulated in Table \ref{TabRes}. Overall $\bar x_{ij}$ factors (see Eq. \eqref{J3form}) are suppressed. On setting all fermions to zero, our results match with the non-supersymmetric case of \cite{Giombi:2011rz}.

This analysis is independent of the free theory computation in the previous subsection but corroborates the parity-even results obtained there, since the parity-even part of the various correlators obtained by applying the conservation constraint is the same as the result of the free theory computation.

\setlength{\tabcolsep}{0pt}
\def\arraystretch{1.2}
\begin{table}
\begin{tabular}{|P{5em}|P{18.5em}|P{17em}|}
\hline
Correlator & Parity-even part& Parity-odd part\\\hline
$\langle J_{0}J_{0}J_{0}\rangle$ & $\vphantom{\bigg.}1+\frac1{16}R'\vphantom{\bigg.} $ & $T'$\\
\hline
\begin{tabular}{c}
     $\langle J_{s}J_{0}J_{0}\rangle$  \\[-.5em]
     $(s\geq2)$ 
\end{tabular}& {
$\vphantom{\bigg.}Q_{1}^{s}\left(
1+\frac{2s+1}{16}R'
\right)\vphantom{\bigg.}$
}&0\\
\hline
$\langle J_{1}J_{1}J_{0}\rangle$ & \begin{tabular}{c}
$Q_{1}Q_{2}\left(
1+\frac{9}{16}R'
\right)+P_{3}^{2}\left(
2+\frac{3}{8}R'
\right)$\\
$+2iP_{3}\bar R_{3}$
\end{tabular} & \begin{tabular}{c}
$Q_{1}Q_{2}T'-S_{3}\bar R_{3}-2iS_{3}P_{3}$
\end{tabular}\\
\hline
\begin{tabular}{c}
     $\langle J_{s}J_{1}J_{0}\rangle$  \\[-.5em]
     $(s\geq2)$ 
\end{tabular} & 
{
\begin{tabular}{c}
$Q_{1}^{s}Q_{2}\left(
1+\frac{3(2s+1)}{16}R'
\right)$\\
$+Q_{1}^{s-1}P_{3}^{2}\left(
2s+\frac{s(2s+1)}{8}R'
\right)$\\
$+2isQ_{1}^{s-1}P_{3}\bar R_{3}$
\end{tabular}
}&0\\
\hline
$\langle J_{2}J_{2}J_{0}\rangle$  &
\begin{tabular}{c}
$Q_{1}^{2}Q_{2}^{2}\left(
1+\frac{25}{16}R'
\right)+Q_{1}Q_{2}P_{3}^{2}\left(
8+\frac{22}{4}R'
\right)$\\
$+8i Q_{1}Q_{2}P_{3}\bar R_{3}+P_{3}^{4}\left(
\frac{8}{3}-\frac12R'
\right)$\\
$+\frac{16i}{3}P_{3}^{3}\bar R_{3}$
\end{tabular}&
\begin{tabular}{c}
$Q_{1}^{2}Q_{2}^{2}T'+Q_{1}Q_{2}P_{3}S_{3}\left(
-\frac{2i}{3}+\frac i6 R'
\right)$\\
$-\frac53 Q_{1}Q_{2}S_{3}\bar R_{3}+P_{3}^{3}S_{3}\left(
-\frac{4i}{3}-\frac i6R'
\right)$
\end{tabular}
\\
\hline
\begin{tabular}{c}
     $\langle J_{s}J_{2}J_{0}\rangle$  \\[-.5em]
     $(s>2)$ 
\end{tabular}  &
{
\begin{tabular}{c}
$Q_{1}^{s}Q_{2}^{2}\left(
1+\frac{5(2s+1)}{16}R'
\right)$\\
$+Q_{1}^{s-1}Q_{2}P_{3}^{2}\left(
4s+\frac{s(6s-1)}{4}R'
\right)$\\
$+Q_{1}^{s-2}P_{3}^{4}\left(
\frac{4s(s-1)}{3}+\frac{s(s-1)(2s-7)}{12}R'
\right)$\\
$+4is Q_{1}^{s-1}Q_{2}P_{3}\bar R_{3}+\frac{8s(s-1)}{3}iQ_{1}^{s-2}P_{3}^{3}\bar R_{3}$
\end{tabular}
}&0
\\
\hline
$\langle J_{2}J_{1}J_{1}\rangle$  &
\begin{tabular}{c}
$Q_{1}^{2}Q_{2}Q_{3}\left(
1+\frac{21}{16}R'
\right)+Q_{1}^{2}P_{1}^{2}\left(
-\frac25-\frac38R'
\right)$\\
$+2iQ_{1}^{2}P_{1}\bar R_{1}+Q_{1}P_{1}P_{2}P_{3}\left(
-\frac85-\frac32R'
\right)$\\
$+P_{2}^{2}P_{3}^{2}\left(
\frac85+\frac12 R'\right)-4iQ_{1}P_{2}P_{3}\bar R_{1}
$
\end{tabular}&
\begin{tabular}{c}
$Q_{1}^{2}Q_{2}Q_{3}T'-Q_{1}^{2}P_{1}S_{1}\left(
2i+2iR'
\right)$\\
$-Q_{1}P_{1}P_{2}P_{3}T'+Q_{1}P_{2}P_{3}S_{1}\left(
4i+2iR'
\right)$\\
$+2Q_1^2S_1\bar R_1-(P_{2}^{2}S_{3}\bar R_{3}+P_{3}^{2}S_{2}\bar R_{2})$
\end{tabular}
\\
\hline
\begin{tabular}{c}
     $\langle J_{s}J_{1}J_{1}\rangle$  \\[-.5em]
     $(s>2)$ 
\end{tabular}  &
{
\begin{tabular}{c}
$Q_{1}^{s}Q_{2}Q_{3}\left(
1+\frac{3(2s+3)}{16}R'
\right)$\\
$+Q_{1}^{s}P_{1}^{2}\left(
-\frac{2(s-1)}{2s+1}-\frac{3(s-1)}8R'
\right)$\\
$ + Q_{1}^{s-1}P_{1}P_{2}P_{3}\left(
-\frac{4s}{2s+1}-\frac{3s}{4} R'
\right)$\\
$+Q_{1}^{s-2}P_{2}^{2}P_{3}^{2}\left(
\frac{4s(s-1)}{2s+1}+\frac{s(s-1)}{4} R'\right)$\\
$+2iQ_{2}^{s}P_{1}\bar R_{1}-2isQ_{1}^{s-1}P_{2}P_{3}\bar R_{1}
$
\end{tabular}
}&0
\\\hline
$\langle J_{2}J_{2}J_{2}\rangle$  &
\begin{tabular}{c}
$Q_{1}^{2}Q_{2}^{2}Q_{3}^{2}\left(
3+\frac{103}{16}R'
\right)$\\
$+P_{1}^{2}P_{2}^{2}P_{3}^{2}\left(
\frac{128}{25}+\frac{72}{5}R'
\right)$\\
$+P_{1}P_{2}P_{3}Q_{1}Q_{2}Q_{3}\left(
-\frac{16}{5}-11R'
\right)$\\
$+\sum_{i}P_{i}^{4}Q_{i}^{2}\left(
-\frac{8}{5}-\frac{5}{2}R'
\right)$\\
$+\sum_{\text{cyc}}P_{1}P_{2}\bar R_{3}\left(
\frac{64i}{5} P_{1}P_{2}P_{3}-8iQ_{1}Q_{2}Q_{3}
\right)$
\end{tabular}&
\begin{tabular}{c}
$9Q_{1}^{2}Q_{2}^{2}Q_{3}^{2}T'+38P_{1}^{2}P_{2}^{2}P_{3}^{2}T'$\\
$-8Q_{1}Q_{2}Q_{3}P_{1}P_{2}P_{3}T'-5T'\sum_{i}P_{i}^{4}Q_{i}^{2}$\\
$+\sum_{i}P_{i}^{3}Q_{i}^{2}S_{i}\left(
-8i-6iR'
\right)$\\
$+P_{1}P_{2}P_{3}\sum_{\text{cyc}}P_{1}P_{2}S_{3}\left(
-40i-46iR'
\right)$
\end{tabular}
\\
\hline
\end{tabular}
\caption{\textbf{Results from conservation constraints}. These results have been checked case by case up to $s=6$. The current superfield $J_0=\bar \Phi \Phi$ has $\Delta=1$.}
\label{TabRes}
\end{table}

The table of results shows clearly that for $\cN=2$ 3d SCFTs the 3-point correlator of conserved currents of various spins always has only two undetermined constant coefficients that come with the two parts - a parity-even part arising from the free theory, and a parity-odd part which comes from interactions:

%\vspace{0.4cm}
\begin{equation}
 \langle J_{s_{1}}J_{s_{2}}J_{s_{3}}\rangle = \frac{1}{{\bar x}_{12}{\bar x}_{23}{\bar x}_{31}}\Big(a \langle J_{s_{1}}J_{s_{2}}J_{s_{3}}\rangle_{even} + b \langle J_{s_{1}}J_{s_{2}}J_{s_{3}}\rangle_{odd}\Big).
\end{equation}
%\vspace{0.4cm}

Furthermore, we also note from the examples, that the coefficient of the parity-odd structure vanishes when the 3-point correlator is `outside the triangle'. \footnote{This means that the inequality between the spins $s_i+s_j \ge s_k$ is violated for some spins, see \cite{Giombi:2011rz}, \cite{Maldacena:2011jn}. Inside the triangle, conservation constraints allow a parity-odd structure, as our table shows. However, for a correlator involving an exactly conserved higher spin ($s>2$) current, we would expect the theory to be free along the lines of \cite{Maldacena:2011jn} and hence the parity-odd structure would not arise,  although this can't be inferred from conservation constraints alone. We would like thank Sachin Jain for discussions regarding this point.}

% table of results

\section{Discussion}\label{dis}

In this paper we studied 3-point correlators of spinning operators in $\cN=2$ 3d SCFTs in terms of superconformal invariants. Our main results include:
\begin{itemize}
    \item Construction of the parity-even invariants and determination of relations between them.
    \item Construction of the parity-odd invariants and determination of relations between them.
    \item Expressions for several 3-point functions of spinning operators in terms of a linearly independent basis made up of the minimal set of invariants.
    \item Computing the 3-point conserved current correlators in the free field case and expressing them in terms of the invariants.
    \item Studying the constraints of conservation - and verifying through several examples that conserved current 3-point correlators contain 2 OPE coefficients comprising of the parity-even and odd parts of the correlator.
\end{itemize}

We would like to emphasise again that the construction of the invariants in the $\cN=2$ case is more involved and subtle compared to the $\cN=1,\,0$ cases due to the presence of R-symmetry. Although the $P,Q,S$  invariants for $\cN=0$ have a relatively straightforward extension in superspace, the fermionic invariants for $\cN=2$ are completely different from the $\cN=1$ case, although built out of similar building blocks which are now charged under R-symmetry. To build our minimal set of $\cN=2$ invariants, we had to deal with two layers of complexity. Firstly, as discussed in Section \ref{PE}, not all R-symmetry singlets built by suitably contracting the covariant building blocks will be super-inversion invariant. Secondly, even when they are, they will not be independent and we found  (in Sections \ref{PE}, \ref{PO} and Appendix \ref{rel}) relations which give our final minimal set of invariants: $P_i,Q_i,\Rb_i, R', S_i, T'$.

It would be interesting to take forward this work in various directions: 
\begin{itemize}
 \item {\it Higher dimensions -- } The same methods should work for correlators in SCFTs in higher than 3 dimensions. The supercovariant  structures in this case are known from the work of Park and Osborn \cite{Park:1997bq, Osborn:1998qu, Park:1998nra, Park:1999pd}. One new feature would be the existence of mixed symmetry tensors not present in the 3d case, which would engender the use of grassmanian polarisation spinors.
    \item {\it More supersymmetry -- } It should be straightforward, though cumbersome, to extend this analysis to higher supersymmetry. Off-shell superspace formulations for SCFTs in three dimensions exist for $\cN=1,2,3,4$ \cite{Gates:1983nr}. For $\cN > 2$, there will be other invariant tensors of $SO(\cN)$ which may also be used to build R-symmetry singlets. For example, for $\cN=3$ we can have $\delta_{ab}$ as well as the anti-symmetric $\epsilon_{abc}$. 
    % Comment on superspace for higher N. 
    \item{\it Generating functions --} Building on the various examples, it would be interesting to work out the form of the generating function for 3-point correlators in the $\cN=1,2$  free theory. The form of this generating function is known for the non-supersymmetric case \cite{Giombi:2011rz} in terms of the parity-even invariants $P\, \text{and}\,\, Q$, but not known for the SCFT case. 
    \item {\it Bootstrap -- } Since 4-point superconformal blocks can be built from the product of 3-point functions of spinning operators, our analysis should aid in implementations of the superconformal bootstrap for 4-point functions in 3d $\cN=2$ SCFTs. Bootstrap studies of 3d SCFTs include \cite{Bobev:2015vsa, Bobev:2015jxa} for $\cN=2$ and \cite{Bashkirov:2013vya, Atanasov:2018kqw, Rong:2018okz} for $\cN=1$ (see also \cite{Chester:2014fya, Chester:2014mea} for higher $\cN$).
    \item {\it Momentum space -- } It would be instructive to carry out our general analysis in momentum space. Firstly, this is useful because most perturbative computations of correlators in specific theories are carried out directly in momentum space. Secondly, the structure of momentum space correlators in (non-supersymmetric) CFTs has been actively investigated in the last decade, starting with the works \cite{Maldacena:2011nz, Coriano:2013jba, Bzowski:2013sza},  with a number of interesting conceptual and technical results and it would be natural to study this also for SCFTs.  In particular, conserved current correlators with spin have a simple form in momentum space (and more so in terms of spinor-helicity variables)  for 3d CFTs \cite{Maldacena:2011nz, Jain:2021wyn, Jain:2021vrv, Jain:2021gwa} and it would be interesting to investigate the same for SCFT correlators.
    \item {$\cN=2$ \it Chern-Simons-Matter theories -- } A prime example of 3d Lagrangian CFTs are the Chern-Simons theories coupled to matter. Such theories are exactly conformal and have a well-controlled weakly broken higher-spin symmetry \cite{Maldacena:2012sf}. Our analysis should be of use in computations of correlators in $\cN =2$ 3d Chern-Simons-Matter theories \cite{Inbasekar:2019wdw}.
\end{itemize}

\section*{Acknowledgements}
AAN would like to thank J.H. Park and S. Jain for useful suggestions and comments on the draft. AJ wishes to acknowledge support from the Council of Scientific and Industrial Research (CSIR) for a Junior Research Fellowship (NET-JRF).

\section*{Appendices}
\appendix
\section{Conventions and useful relations}\label{conv}

We use the mostly plus $\{-1,1,1\}$ metric convention. Spinors - which can be taken to be Majorana  - transform under the (double cover of the) 3d Lorentz group, which is $SL(2, R)$.
We thus use the fundamental representation and impose the Majorana condition on the 2-component spinors which is the following reality condition $\psi_\alpha = \psi^*_\alpha$. In this representation, superconformal theories with $\cN$ extended supersymmetry possess an $SO(\cN)$ R-symmetry group which is a subgroup of the full superconformal group.
For 3 dimensions, we can make a choice of a real basis for the $\gamma$ matrices
\be
(\gamma_\mu)_\alpha^{~\beta} \equiv (i\sigma^2,\sigma^1,\sigma^3) = \left(\left(\begin{array}{cc} \mbox{  }0 & \mbox{  }1  \\ -1 & \mbox{ }0 \end{array}\right),
\left(\begin{array}{cc} 0 & 1  \\ \mbox{ }1 & \mbox{ }0 \end{array}\right),\left(\begin{array}{cc} 1 & \mbox{    }0  \\ 0 & -1 \end{array}\right)\right)
\ee
These matrices when both indices are up (or down) are symmetric
\begin{equation}
\begin{split}
 (\gamma_\mu)_{\alpha\beta} \equiv (\mathds{1},\sigma^3,-\sigma^1) &~~~~~~~ (\gamma_\mu)^{\alpha\beta} \equiv (\mathds{1},-\sigma^3,\sigma^1)
\end{split}
\end{equation}
The anti-symmetric $\epsilon$ symbol is defined as $\epsilon^{12} = -1 = \epsilon_{21}$.
\be
\begin{split}
\epsilon \gamma^\mu \epsilon^{-1} = -(\gamma^\mu)^T\\
\epsilon \Sigma^{\mu\nu} \epsilon^{-1} = -(\Sigma^{\mu\nu})^T
\end{split}
\ee
with $\Sigma^{\mu\nu} = -\frac{i}{4}[\gamma^\mu,\gamma^\nu]$ as the Lorentz group generators.  
The $\gamma$ matrices satisfy
\be
(\gamma_\mu\gamma_\nu)_\alpha^{~\beta} =  \eta_{\mu\nu}\delta_{\alpha}^{~\beta} + \epsilon_{\mu\nu\rho}(\gamma^\rho)_\alpha^{~\beta}
\ee
with $\epsilon_{\mu\nu\rho}$ as the Levi-Civita symbol, and we use $\epsilon_{012} = 1$ ($\epsilon^{012} = -1$).

The spinor Lorentz transformation is
$$\psi^\prime_\alpha \rightarrow -(\Sigma_{\mu\nu})_\alpha^{~\beta}\psi_\beta.$$

The conventions for raising and lowering with $\epsilon$ are
\be
\begin{split}
\psi^\beta = \epsilon^{\beta\alpha}\psi_\alpha\\
\psi_\alpha = \epsilon_{\alpha\beta}\psi^\beta
\end{split}
\ee
and we denote $\psi\chi \equiv \psi^\alpha\chi_\alpha$.
%which results in a sign under Hermitian conjugation
%$$(\psi\chi)^* = -\chi^*\psi^*.$$

\subsection{Superconformal generators and algebra}
The generators of the 3d superconformal algebra in differential operator form are:
\begin{equation}
\begin{split}
P_\mu &= -i \partial_\mu, \\
M_{\mu\nu} &= -i\left(x_{\mu}\partial_{\nu}- x_\nu \partial_\mu - 
               \frac{1}{2}\epsilon_{\mu\nu\rho}(\gamma^{\rho})_\alpha^{\mbox{ }\beta}\theta^a_\beta \frac{\partial}{\partial\theta^a_{\alpha}}\right) + {\cal M}_{\mu\nu}, \\
D &= -i\left(x^\nu\partial_\nu +\frac{1}{2}\theta^{\alpha a}\frac{\partial}{\partial\theta^\alpha_a}\right) 
      + \Delta, \\
% K_\mu &=  x^\nu M_{\nu\mu} - x_\mu D,  \\
K_\mu &= -i\left(\left(x^2 + \frac{(\theta^a\theta^a)^2}{16}\right)\partial_\mu -2 x_\mu\left(x\cdot\partial + 
          \theta^{\alpha a}\frac{\partial}{\partial\theta^\alpha_a}\right)+(\theta^a X_+ \gamma_\mu)^\beta\frac{\partial}{\partial\theta_a^{\beta}}\right) \\
 &= x^\nu M_{\nu\mu} - x_\mu D + \frac{i}2 (\theta^a\gamma_\mu X)^\alpha\frac{\partial}{\partial\theta_a^\alpha} -
        \frac{i}{16}(\theta^a\theta^a)^2\partial_\mu + \frac{(\theta^a\theta^a)}{4}(\theta^b\gamma_\mu)^\alpha\frac{\partial}{\partial\theta_b^\alpha},\\
I^{ab}   &= -i\left(\theta^a \frac{\partial}{\partial\theta_b} - \theta^b\frac{\partial}{\partial\theta_a}\right) + {\cal I}^{ab}, \\
Q_\alpha^a &= \frac{\partial}{\partial\theta^\alpha_a} - \frac{i}{2}\theta^{\beta a}(\gamma^\mu)_{\beta\alpha}\partial_\mu, \\
S_\alpha^a &= -(X_+)_\alpha^{~\beta}Q_\beta^a - i \theta^a \theta^b \frac{\partial}{\partial\theta^\alpha_b} - i \theta^{a}_\alpha \theta^{b\beta}\frac{\partial}{\partial\theta^\beta_b}
                +\frac{i}{2}(\theta^b\theta^b)\frac{\partial}{\partial\theta_a^\alpha} \\
           &= -(X_-)_\alpha^{~\beta}\frac{\partial}{\partial\theta_a^\beta} + \frac{\theta_\alpha^a}{2} D + \frac{1}{4}\epsilon_{\mu\nu\rho}(\gamma^\rho\theta^a)_\alpha M^{\mu\nu}
              -\frac{(\theta^b\theta^b)}{8}\theta^{a\beta} \del_{\beta\alpha} - \frac{3i}{4}\left(\theta_\alpha^a\theta\frac{\partial}{\partial\theta} +
               \theta^a\theta^b\frac{\partial}{\partial\theta_b^\alpha}\right)~.
% -x_\alpha^{~\beta} \frac{\partial}{\partial \theta^\beta}- 
%             \frac{i}{2}\theta_\alpha x^\nu \partial_\nu- 
%             \frac{i}{4}\theta\theta\frac{\partial}{\partial\theta^\alpha}+
%             \frac{i}{2}(\gamma_\mu)_\alpha^\beta\theta_\beta \epsilon^{\mu\nu\rho}x_{\nu}\partial_{\rho}  ~.
\end{split}
\label{difalg}
\end{equation}
where $\cal{M}, \cal{I}$ arise due to the tensor structure of the operators on which these generators act.

The superconformal algebra generated by these is:
\begin{equation}
\begin{split}
[ M_{\mu\nu}, M_{\rho\lambda} ]&= i\left( \eta_{\mu\rho}M_{\nu\lambda}- \eta_{\nu\rho}M_{\mu\lambda}- 
                                          \eta_{\mu\lambda}M_{\nu\rho}+ \eta_{\nu\lambda}M_{\mu\rho} \right),   \\
[ M_{\mu\nu}, P_\lambda ]&= i(\eta_{\mu\lambda}P_\nu- \eta_{\nu\lambda}P_\mu),  \\
[ M_{\mu\nu}, K_\lambda ]&= i(\eta_{\mu\lambda}K_\nu- \eta_{\nu\lambda}K_\mu),  \\
[ D, P_\mu ]&= iP_{\mu} ~,~~~ [ D, K_\mu ]= -iK_\mu, \\
[ P_\mu, K_\nu ]&= 2i(\eta_{\mu\nu} D- M_{\mu\nu}),  \\
[ I_{ab}, I_{cd} ] &= i\left(\delta_{ac}I_{bd} - \delta_{bc}I_{ad} - \delta_{ad}I_{bc} + \delta_{bd}I_{ac}\right),\\
% [ I_{ab}, M_{\mu\nu} ] &= [ I_{ab}, P_\mu]=0, \\
\{ Q_\alpha^a, Q_\beta^b \}&= (\gamma^{\mu})_{\alpha\beta}P_\mu\delta^{ab}, \\
% \{ D_\alpha, D_\beta \}&= -(\gamma^{\mu})_{\alpha\beta}P_\mu, \\
[I_{ab}, Q^\alpha_c] &= i(\delta_{ac}Q^\alpha_b - \delta_{bc}Q^\alpha_a) , \\
\{ S_\alpha^a, S_\beta^b \}&= (\gamma^{\mu})_{\alpha\beta}K_\mu\delta^{ab}, \\
[I_{ab}, S^\alpha_c] &= i(\delta_{ac}S^\alpha_b - \delta_{bc}S^\alpha_a) , \\
[ K_\mu, Q_\alpha^a ]&= i(\gamma_\mu)_{\alpha}^{~\beta}S_{\beta}^a, \\
[ P_\mu, S_\alpha^a ]&= i(\gamma_\mu)_{\alpha}^{~\beta}Q_{\beta}^a, \\
[ D, Q_\alpha^a ]&= \frac{i}{2} Q_\alpha^a ~,~~~ [ D, S_\alpha^a ]= -\frac{i}{2} S_\alpha^a, \\
[M_{\mu\nu}, Q_\alpha^a]&= -(\Sigma_{\mu\nu})_\alpha^{~\beta}Q_\beta^a, \\
[M_{\mu\nu}, S_\alpha^a]&= -(\Sigma_{\mu\nu})_\alpha^{~\beta}S_\beta^a, \\
\{ Q^a_\alpha, S^b_\beta \}&= \left(\epsilon_{\beta\alpha}D - \half\epsilon_{\mu\nu\rho}(\gamma^\rho)_{\alpha\beta}M^{\mu\nu}\right)\delta^{ab} + \epsilon_{\beta\alpha}I^{ab}.
\end{split}
\label{N1lag}
\end{equation}
The rest of the (anti)-commutators are zero. 

%change text: Here the derivative expressions act on superspace coordinates while the operators $\cal M$, $\Delta$ and ${\cal I}^{ab}$ act on the operators (states) which carry tensor structure, non-zero scaling dimensions and transform non-trivially under $R$-symmetry. All indices are contracted in matrix notation (the spinors are contracted from north-west to south-east)

\section{More relations between invariants}\label{rel}
We had the following set of superconformal invariants
which were used to construct various 3-point correlators:
\begin{alignat}{2}
&\text{parity even:}\quad &&P_{i},\  Q_{i},\  \bar R_{i},\ R'\\
&\text{parity odd:}\quad &&S_{i},\  T'
\end{alignat}
We noted earlier that using the covariant structures we seem to be able to build other invariants as well. However, all such additional invariants can  be expressed in terms of the above minimal set of invariants. We show this explicitly below by listing manifold relations where an invariant structure is expressed in terms of our minimal set. We have classified the relations according to the degree in $\lambda$ and $\T$, an ${\cal O}(\lambda^m \Theta^{n})$ relation will have terms, each of which has $m$ $\lambda$'s and $n$ $\Theta$'s.

\subsection{Parity even relations}
In this subsection we enumerate parity-even invariant structures built from the covariant building blocks. These will all be expressed in terms of our claimed minimal set of parity-even invariants: $P_i, Q_i, \Rb_i, R'$.
\begin{enumerate}

	\item {${\cal O}(\Theta^{2})$:}
	\begin{enumerate}
		
		\item $\displaystyle \frac{\bar x_{12}^{2}\bar x_{31}^{2}}{\bar x_{23}^{2}}\Omega^{aa}_{1}=\frac i2 R'$
	
	\end{enumerate}
	
	\item {${\cal O}(\Theta^{4})$:}
	\begin{enumerate}

		\item $\bar x_{12}^{4}\Sigma_{12}^{ab}\Sigma_{12}^{ab}=-R',\quad \bar{x}_{12}^{4}\Omega_{1}^{ab}\Omega_{2}^{ab}=R'$\\[.2em]
			
	\end{enumerate}

	\item {${\cal O}(\lambda^{2} \Theta^{2})$:}
	\begin{enumerate}
	
		\item $R_{1}^{a}R_{1}^{a}=\dfrac{\bar x_{23}^{2}}{\bar x_{12}^{2}\bar x_{31}^{2}}\pi_{12}^{a}\pi_{12}^{a}=\dfrac{\bar x_{23}^{2}\bar x_{12}^{2}}{\bar x_{31}^{2}}\sigma_{12}^{a}\sigma_{12}^{a}=\dfrac{\bar x_{12}^{2}\bar x_{31}^{2}}{\bar x_{23}^{2}}\omega_{1}^{a}\omega_{1}^{a}=0$\\[.2em]
	
		\item $\sqrt{\dfrac{\bar x_{12}^{2}\bar x_{31}^{2}}{\bar x_{23}^{2}}}R_{1}^{a}\sigma_{21}^{a}=-\bar R_{3},\quad \dfrac{\bar x_{23}^{2}\bar x_{31}^{2}}{\bar x_{12}^{2}}\sigma_{13}^{a}\sigma_{23}^{a}=\bar R_{3}-\dfrac i2 R' P_{3},\quad \dfrac{\bar x_{23}^{2}}{\bar x_{13}^{2}}\pi_{12}^{a}\omega_{2}^{a}=\bar R_{3}$\\[.2em]
	
	\end{enumerate}
	
	\item {${\cal O}(\lambda^{2} \Theta^{4})$:}
	\begin{enumerate}
		
		\item $R_{1}^{a}\Pi_{23}^{ab}R_{2}^{b}=-\dfrac12 R' P_{3},\quad \dfrac{\bar x_{12}^{2}\bar x_{31}^{2}}{\bar x_{23}^{2}}R_{1}^{a}\Omega_{1}^{ab}R_{1}^{b}=-\dfrac12 R' Q_{1}$\\[.2em]
		
		\item $\dfrac{1}{\bar x_{12}^{2}}\pi_{12}^{a}\Pi_{23}^{ab}\pi_{23}^{b}=\dfrac12 R' P_{3},\quad \bar{x}_{23}^{2}\sigma_{12}^{a}\Pi_{23}^{ab}\sigma_{23}^{b}=-\dfrac12 R' P_{3},\quad \bar{x}_{12}^{2}\omega_{1}^{a}\Pi_{12}^{ab}\omega_{2}^{b}=-\dfrac12 R' P_{3},\\ \pi_{12}^{a}\Pi_{21}^{ab}\omega_{1}^{b}=-\dfrac12 R' Q_{1}$\\[.2em]
		
		\item ${\bar{x}_{12}^{2}}\pi_{12}^{a}\Sigma_{21}^{ab}\sigma_{31}^{b}=-\dfrac12 R' P_{2},\quad {\bar{x}_{23}^{4}}\sigma_{12}^{a}\Sigma_{23}^{ab}\omega_{3}^{b}=\dfrac12 R' P_{2}$\\[.2em]
		
		\item $\dfrac{\bar x_{12}^{2}\bar x_{23}^{2}}{\bar x_{13}^{4}}\pi_{12}^{a}\Omega_{2}^{ab}\pi_{32}^{b}=-\dfrac12 R' P_{2},\quad  \dfrac{\bar x_{23}^{4}\bar x_{12}^{2}}{\bar x_{13}^{4}}\pi_{12}^{a}\Omega_{2}^{ab}\omega_{2}^{b}=\dfrac12 R' P_{3},\quad \\ \dfrac{\bar x_{12}^{4}\bar x_{31}^{4}}{\bar x_{23}^{4}}\omega_{1}^{a}\Omega_{1}^{ab}\omega_{1}^{b}=\dfrac12 R' Q_{1},\quad \dfrac{\bar x_{12}^{4}\bar x_{31}^{4}}{\bar x_{23}^{4}}\sigma_{21}^{a}\Omega_{1}^{ab}\sigma_{31}^{b}=-\dfrac12 R' P_{1}$\\[.2em]
	
	\end{enumerate}
	
	\item {${\cal O}(\lambda^{4} \Theta^{4})$:}
	\begin{enumerate}
	
		\item $(R_{1}^{a}R_{2}^{a})^{2}=-\dfrac12 R'(Q_{1}Q_{2}-P_{3}^{2})$\\[.2em]
		
		\item $\dfrac{1}{\bar x_{12}^{4}}(\pi_{12}^{a}\pi_{23}^{a})^{2}=-\dfrac12 R'(Q_{1}Q_{2}-P_{3}^{2}),\quad \dfrac{1}{\bar x_{12}^{4}}(\pi_{12}^{a}\pi_{21}^{a})^{2}=-\dfrac12 R'(Q_{1}Q_{2}-P_{3}^{2}),\\
		\dfrac{1}{\bar x_{13}^{4}}(\pi_{12}^{a}\pi_{31}^{a})^{2}=-\dfrac12 R'(Q_{1}Q_{3}-P_{2}^{2})$\\[.2em]
		
		\item $\bar x_{23}^{4}(\sigma_{12}^{a}\sigma_{23}^{a})^{2}=-\dfrac12 R'(Q_{1}Q_{2}-P_{3}^{2}),\quad \bar x_{12}^{4}(\sigma_{12}^{a}\sigma_{21}^{a})^{2}=-\dfrac12 R'(Q_{1}Q_{2}-P_{3}^{2}),\\
		 \bar x_{12}^{4}(\sigma_{12}^{a}\sigma_{31}^{a})^{2}=-\dfrac12 R'(Q_{1}Q_{3}-P_{2}^{2})$\\[.2em]
		
		\item $\bar x_{12}^{4}(\omega_{1}^{a}\omega_{2}^{a})^{3}=-\dfrac12 R'(Q_{1}Q_{2}-P_{3}^{2})$\\[.2em]
		
		\item $\dfrac{\bar x_{31}^{2}}{\bar x_{23}^{2}\bar x_{12}^{2}}(R^{a}_{1}\pi^{a}_{21})^{2}=\dfrac12 R'P_{3}^{2},\quad \dfrac{\bar x_{31}^{2}}{\bar x_{23}^{2}\bar x_{12}^{2}}(R^{a}_{1}\pi^{a}_{23})^{2}=\dfrac12 R'P_{3}^{2},\\ \quad \dfrac{\bar x_{23}^{2}}{\bar x_{12}^{2}\bar x_{31}^{2}}(R^{a}_{1}\pi^{a}_{12})^{2}=\dfrac12 R'Q_{1}^{2},\\
		\dfrac{1}{\bar x_{13}^{2}}R_{1}^{a}\pi_{12}^{a}R_{1}^{b}\pi^{b}_{32}=-\dfrac12 Q_{1}P_{2},\quad \dfrac{1}{\bar x_{23}^{2}}R_{1}^{a}\pi_{21}^{a}R_{1}^{b}\pi^{b}_{31}=-\dfrac12 R'P_{2}P_{3},\\
		\dfrac{1}{\bar x_{12}^{2}}R_{1}^{a}\pi_{23}^{a}R_{1}^{b}\pi^{b}_{13}=-\dfrac12 R'Q_{1}P_{3}$\\[.2em]
		
		\item $\dfrac{\bar x_{13}^{2}\bar x_{12}^{2}}{\bar x_{23}^{2}}(R^{a}_{1}\sigma^{a}_{21})^{2}=-\dfrac12 R'(Q_{1}Q_{2}-P_{3}^{2}),\quad \dfrac{\bar x_{23}^{2}\bar x_{13}^{2}}{\bar x_{12}^{2}}(R^{a}_{1}\sigma^{a}_{23})^{2}=-\dfrac12 R'(Q_{1}Q_{2}-P_{3}^{2}),\quad \\
		\dfrac{\bar x_{23}^{2}\bar x_{12}^{2}}{\bar x_{31}^{2}}(R^{a}_{1}\sigma^{a}_{12})^{2}=0$\\[.2em]
		
		\item $\dfrac{\bar x_{13}^{2}\bar x_{12}^{2}}{\bar x_{23}^{2}}(R^{a}_{1}\omega^{a}_{1})^{2}=\dfrac12 R'Q_{1}^{2},\quad \dfrac{\bar x_{12}^{2}\bar x_{23}^{2}}{\bar x_{13}^{2}}(R^{a}_{1}\omega^{a}_{2})^{2}=\dfrac12 R'P_{3}^{2}$\\[.2em]
		
		\item $(\pi_{12}^{a}\sigma_{21}^{a})^{2}=\dfrac12 R'P_{3}^{2},\quad (\pi_{12}^{a}\sigma_{31}^{a})^{2}=\dfrac12 R'P_{2}^{2},\quad \dfrac{\bar x_{23}^{4}}{\bar x_{13}^{4}}(\pi_{12}^{a}\sigma_{21}^{a})^{2}=\dfrac12 R'P_{2}^{2},\\
		 \dfrac{\bar x_{23}^{4}}{\bar x_{12}^{4}}(\pi_{12}^{a}\sigma_{21}^{a})^{2}=\dfrac12 R'P_{3}^{2}$
			
	\end{enumerate}

\end{enumerate}
Note that other relations can be generated by cyclic permutation of $1,2,3$ in the above list.

\subsection{Parity odd relations} 
In this subsection we list the various parity-odd invariant structures built from the covariant building blocks. These will all be expressed in terms of our minimal set of invariants.
\begin{enumerate}
	
	\item {${\cal O}(\Theta^{4})$}:
	
	\begin{enumerate}
	 
		\item $\bar{x}^{2}_{12}\Pi^{ab}_{12}\Sigma^{ab}_{12}=0$\\[.2em]
	
	\end{enumerate}

\end{enumerate}

\begin{enumerate}

	\item {${\cal O}(\lambda^{2} \Theta^{2})$}:
	
	\begin{enumerate}
	
		\item $\sqrt{\dfrac{\bar x_{12}^{2}\bar x_{31}^{2}}{\bar x_{23}^{2}}}R_{1}^{a}\omega_{1}^{a}=\dfrac12 T' Q_{1},\quad \sqrt{\dfrac{\bar x_{13}^{2}}{\bar x_{12}^{2}\bar x_{23}^{2}}}R_{1}^{a}\pi_{21}^{a}=\dfrac12 T'P_{3}$\\[.2em]	
		
		\item $\dfrac{\bar x_{23}^{2}}{\bar x_{13}^{2}}\pi_{12}^{a}\sigma_{12}^{a}=\dfrac12 T' Q_{1},\quad \dfrac{\bar x_{23}^{2}}{\bar x_{13}^{2}}\pi_{12}^{a}\sigma_{32}^{a}=\dfrac12 T' P_{2},\quad {\dfrac{\bar x_{13}^{2}}{\bar x_{23}^{2}\bar x_{23}^{2}}}\sigma_{12}^{a}\omega_{2}^{a}=\dfrac12 T' P_{3}-\dfrac i2 R' S_{3}$\\[.2em]
	
	\end{enumerate}
	
	\item {${\cal O}(\lambda^{2} \Theta^{4})$}:

	\begin{enumerate}
	
		\item $\bar x_{12}^{2}R_{1}^{a}\Sigma_{12}^{ab}R_{2}^{b}=-\dfrac12 R' S_{3},\quad \bar x_{12}^{4}\omega_{1}^{a}\Sigma_{12}^{ab}\omega_{2}^{b}=-\dfrac12 R' S_{3},\quad \pi_{21}^{a}\Sigma_{12}^{ab}\pi_{12}^{b}=\dfrac12 R' S_{3},\\
		\dfrac{\bar x_{23}^{2}\bar x_{13}^{2}}{\bar x_{12}^{4}}\pi_{31}^{a}\Sigma_{12}^{ab}\pi_{32}^{b}=0,\quad \dfrac{\bar x_{12}^{2}}{\bar x_{23}^{4}}\pi_{21}^{a}\Sigma_{12}^{ab}\pi_{32}^{b}=-\dfrac12 R'S_{1},\quad \dfrac{\bar x_{12}^{2}}{\bar x_{13}^{4}}\pi_{31}^{a}\Sigma_{12}^{ab}\pi_{12}^{b}=-\dfrac12 R'S_{2}$\\[.2em]
		
		\item $\sqrt{\dfrac{\bar x_{23}^2}{\bar x_{13}^2\bar x_{12}^2}}R_{1}^{a}\Pi_{12}^{ab}\pi_{12}^{b}=0,\quad \sqrt{\dfrac{\bar x_{12}^2}{\bar x_{13}^2\bar x_{32}^2}}R_{1}^{a}\Pi_{12}^{ab}\pi_{32}^{b}=\dfrac12 R'S_{2},\\
		 \sqrt{\dfrac{\bar x_{12}^2\bar x_{23}^2}{\bar x_{13}^2}}R_{1}^{a}\Pi_{12}^{ab}\omega_{2}^{b}=-\dfrac12 R' S_{3},\quad \pi_{21}^{a}\Pi_{12}^{ab}\sigma_{12}^{b}=\dfrac12 R'S_{3},\quad \pi_{21}^{a}\Pi_{12}^{ab}\sigma_{32}^{b}=-\dfrac12 R'S_{1},\\
		  \dfrac{\bar x_{12}^2}{\bar x_{13}^2}\pi_{31}^{a}\Pi_{12}^{ab}\sigma_{12}^{b}=\dfrac12 R'S_{3},\quad \bar x_{12}^2\omega_{1}^{a}\Pi_{12}^{ab}\sigma_{12}^{a}=0$\\[.2em]
		
		\item ${\dfrac{\bar x_{12}^2\bar x_{13}^4}{\bar x_{23}^4}}\pi_{21}^{a}\Omega_{1}^{ab}\sigma_{21}=0,\quad {\dfrac{\bar x_{13}^2\bar x_{12}^4}{\bar x_{23}^4}}\pi_{31}^{a}\Omega_{1}^{ab}\sigma_{21}=-\dfrac12 R'S_{1},\quad \dfrac{\bar x_{12}^4\bar x_{13}^4}{\bar x_{23}^4}\omega_{1}^{a}\Omega_{1}^{ab}\sigma_{21}^{a}=\dfrac12 R'S_{3}$\\[.2em]
			
	\end{enumerate}
	
	\item {${\cal O}(\lambda^{4} \Theta^{4})$}:

	\begin{enumerate}
	
		\item $\sqrt{\dfrac{\bar x_{23}^{2}}{\bar x_{12}^{2}\bar x_{31}^{2}}}R_{1}^{a}R_{2}^{a}R_{1}^{b}\pi_{12}^{b}=\dfrac12 R' S_{3}Q_{1},\quad \sqrt{\dfrac{\bar x_{12}^{2}}{\bar x_{23}^{2}\bar x	_{31}^{2}}}R_{1}^{a}R_{2}^{a}R_{1}^{b}\pi_{32}^{b}=-\dfrac12 R'S_{3}P_{2},\\ 
		\sqrt{\dfrac{\bar x_{12}^{2}\bar x_{23}^{2}}{\bar x_{31}^{2}}}R_{1}^{a}R_{2}^{a}R_{1}^{b}\omega_{2}^{b}=\dfrac12 R' S_{3}P_{3}$\\[.2em]
		
		\item $\dfrac{1}{\bar x_{12}^2}\pi_{21}^{a}\pi_{12}^{a}\pi_{21}^{b}\sigma_{12}^{b}=-\dfrac12 R'S_{3}P_{3},\quad \dfrac{\bar x_{ik}^4}{\bar x_{23}^4}\pi_{21}^{a}\sigma_{21}^{a}\pi_{21}^{b}\omega_{1}^{b}=-\dfrac12 R'Q_{2}S_{3}$\\[.2em]
	
	\end{enumerate}
	
\end{enumerate}
Again, we can obtain other relations by cyclically permuting $1,2,3$ above.

\subsection{A few more relations}

Using the two relations \eqref{GPE} and \eqref{FPE}, other relations with higher homogeneity in $\lambda's$ can be generated. For example, using Eq. \eqref{FPE}, one gets the  ${\cal O}(\lambda_{1}^{2}\lambda_{2}^2\lambda_{3}^2)$ fermionic relation
\be
P_{1}P_2\bar R_{3}+P_{2}P_3\bar R_{1}+P_3P_1\Rb_2+\frac{1}{2}\sum_i Q_{i}P_i\bar R_{i}-\frac{3i}{4} R'P_{1}P_{2}P_3=0.
\ee

More combinations of the relations yield
\begin{align}
    \begin{aligned}
&\frac{i}{2}\sum_{i}P_{i}^{3}Q_{i}^{2}\bar R_{i}+\frac{i}{2}\sum_{\text{cyc}}P_{1}^{2}Q_{1}(P_{2}Q_{2}\bar R_{2}+P_{3}Q_{3}\bar R_{3})+2\sum_{\text{cyc}}P_{1}^{2}P_{2}^{2}Q_{1}Q_{2}+\sum_{i}P_{i}^{4}Q_{i}^{2}\\
&\qquad\qquad-Q_{1}Q_{2}Q_{3}\sum_{i}P_{i}^{2}Q_{i}+\frac{1}{4}R'P_{1}P_{2}P_{3}\sum_{i}P_{i}^{2}Q_{i}-2P_{1}P_{2}P_{3}\sum_{i}P_{i}^{2}Q_{i}=0
\end{aligned}\\
\begin{aligned}
&-\frac{i}{4}\sum P_{i}^{3}Q_{i}^{2}\bar R_{i}-\frac{1}{2}\sum_{\text{cyc}}P_{1}^{2}P_{2}^{2}Q_{1}Q_{2}+\frac14 Q_{1}Q_{2}Q_{3}\sum_{i}P_{i}^{2}Q_{i}-\sum_{i}P_{i}^{4}Q_{i}^{2}+P_{1}^{2}P_{2}^{2}P_{3}^{2}\\
&\qquad\qquad\qquad+\frac{1}{2}Q_{1}Q_{2}Q_{3}P_{1}P_{2}P_{3}-\frac{1}{8}P_{1}^{2}P_{2}^{2}P_{3}^{2}R'=0
\end{aligned}
\end{align}

Also, we can have purely fermionic relations, since the product of $R'$ with any other fermionic invariant vanishes,
\begin{gather}
\left(
2P_{1}P_{2}P_{3}-\sum_{i}P_{i}^{2}Q_{i}+Q_{1}Q_{2}Q_{3}
\right)R'=0\\
\left(
2\sum_{\text{cyc}}P_{1}^{2}P_{2}^{2}Q_{1}Q_{2}-Q_{1}Q_{2}Q_{3}\sum_{i}P_{i}^{2}Q_{i}-2P_{1}P_{2}P_{3}\sum_{i}P_{i}^{2}Q_{i}+\sum_{i}P_{i}^{4}Q_{i}^{2}
\right)R'=0
\end{gather}
and so on.

Higher order parity-odd relations can also be generated. For example, at ${\cal O}(\lambda_{1}^{4}\lambda_{2}^{4}\lambda_{3}^{4})$ we have:
	\begin{gather}
	\left(
	\frac{1}{2}\sum_{\text{cyc}}Q_{1}P_{1}^{2}(P_{2}Q_{2}S_{2}+P_{3}Q_{3}S_{3})-\frac{1}{2}\sum_{i}P_{i}^{3}Q_{i}^{2}S_{i}+P_{1}P_{2}P_{3}\sum_{i}P_{i}Q_{i}S_{i}
	\right)R'=0
	\end{gather}

These relations are utilised in Section \ref{3pt} where we write the linearly independent parity-even structures for various 3-point functions.

\newpage
\section{Calculational details for free theory correlators}\label{FT}
In this appendix we provide some calculation details for results quoted in Section \ref{CC}. In particular, we will explain through two examples how one computes current correlators in the free-field theory using Wick contractions of the chiral scalar superfield.
For ${\cal N}=2$, the basic superfields are (anti-)chiral,
\begin{align}
\bar D_{\alpha}\Phi=D_{\alpha}\bar \Phi=0
\end{align}
and obey the following free field equations of motion,
\begin{align}
D^{\alpha}D_{\alpha}\Phi=\bar D^{\alpha}\bar D_{\alpha}\bar \Phi=0\,\,.
\end{align}
Solving the above, we get the super-multiplet for the (anti-)chiral superfields to have the form
\begin{align}
\Phi&=\phi+\theta\psi-\frac i2 \theta \gamma^{\mu}\bar\theta\partial_{\mu}\phi\\
\bar\Phi&=\bar\phi+\bar\theta\psi^{*}+\frac i2 \theta \gamma^{\mu}\bar\theta\partial_{\mu}\phi\,\,.
\end{align}
Here $\phi$ and $\psi$ satisfy, respectively, the free Klein-Gordon and Dirac equations, ($\theta\gamma^{\mu}\bar\theta \equiv \theta^{\alpha}(\gamma^{\mu})_{\alpha}^{\ \beta}\bar \theta_{\beta}$).

Next, one can construct the 2-point function
\begin{align}
\langle\bar\Phi_{1}\Phi_{2}\rangle=\frac{1}{y_{12}}
\end{align}
where $\Phi_{i}$ is the superfield at superspace location $(x_{i},
\theta_i)$, while $y_{12}$ is built from $\bar x_{12}$ \cite{Buchbinder:2015qsa},
\begin{align}
y_{12}^{2}=\bar x_{12}^{2}v_{12},\quad v_{12}=\frac{1}{2}\left((V_{12})^{aa}+i\epsilon_{ab}(V_{12})^{ab}\right)
\end{align}
where we've used the 2-point structure $V_{12}^{ab}=\delta^{ab}+i\theta_{12}^{a\alpha}(X_{12+}^{-1})_{\alpha}^{\ \beta}\theta_{12\beta}^{b}$.

This form of the 2-point correlator can be explicitly checked by expanding the superfields and using the expressions
\begin{align}
\langle
\bar\phi_{1}\phi_{2}
\rangle=
\langle
\phi_{1}\bar\phi_{2}
\rangle&=\frac{1}{x_{12}}\\
\langle
\psi_{1\alpha}^{*}\psi_{2\beta}
\rangle=-\langle
\psi_{2\alpha}\psi_{1\beta}^{*}
\rangle&=i\frac{(X_{12})_{\alpha\beta}}{x_{12}^{3}}
\end{align}

\subsection{Computation of $\langle J_{0}J_{0}J_{0}\rangle$}

The spin-0 conserved supercurrent is
\begin{align}
J_{0}=\bar\Phi \Phi
\end{align}
The correlator $\langle J_{0}J_{0}J_{0}\rangle$ can be written as
\begin{align}
\langle J_{0}J_{0}J_{0}\rangle=\left\langle
(\bar\Phi_{1} \Phi_{1})(\bar\Phi_{2} \Phi_{2})(\bar\Phi_{3} \Phi_{3})
\right\rangle
\end{align}
%where the subscripts on the right side denote the superspace location of the fields.

Since this is a free theory correlator it can be computed by Wick contractions on the the right hand side,
\begin{align}
\begin{aligned}
\left\langle
(\bar\Phi_{1} \Phi_{1})(\bar\Phi_{2} \Phi_{2})(\bar\Phi_{3} \Phi_{3})
\right\rangle&=\langle
\wick[offset=1em,sep=.5em]{
\c3{\bar\Phi_{1}} \c1{\Phi_{1}}\,\c1{\bar\Phi_{2}}\c2{ \Phi_{2}}\,\c2{\bar\Phi_{3}} \c3{\Phi_{3}
}}\rangle+
\langle
\wick[offset=1em,sep=.5em]{
\c1{\bar\Phi_{1}} \c2{\Phi_{1}}\,\c3{\bar\Phi_{2}}\c1{ \Phi_{2}}\,\c2{\bar\Phi_{3}} \c3{\Phi_{3}
}}\rangle\\
&=\left(
\frac{1}{y_{13}}
\right)
\left(
\frac{1}{y_{21}}
\right)
\left(
\frac{1}{y_{32}}
\right)+\left(
\frac{1}{y_{12}}
\right)
\left(
\frac{1}{y_{31}}
\right)
\left(
\frac{1}{y_{23}}
\right)
\end{aligned}
\end{align}

We find that the result of this computation is equal to a linear combination of the parity even invariant structures. Since there is no spin in this example, the only one allowed is $R'$,
\begin{align}
\langle J_{0}J_{0}J_{0}\rangle=
\frac{-2}{\bar x_{12}\bar x_{23}\bar x_{31}}
\left(
1+\frac{1}{16}R'
\right)
\end{align}

\subsection{Computation of $\langle J_{1}J_{1}J_{0}\rangle$}

The spin-0 and spin-1 conserved supercurrents are,
\begin{align}
J_{0}=\bar\Phi \Phi\ ,\qquad J_{1}=\bar\Phi(\partial\Phi)-2\bar D\bar\Phi D \Phi-(\partial\bar\Phi)\Phi
\end{align}
Thus, the 3-point correlator
\begin{align}
\begin{aligned}
\langle J_{1}J_{1}J_{0}\rangle&=
\left\langle
\left(
\bar\Phi_{1}(\partial_{1}\Phi_{1})-2\bar D_{1}\bar\Phi_{1} D_{1} \Phi_{1}-(\partial_{1}\bar\Phi_{1})\Phi_{1}
\right)\right.\\
&\left.\qquad\qquad\times
\left(
\bar\Phi_{2}(\partial_{2}\Phi_{2})-2\bar D_{2}\bar\Phi_{2} D_{2} \Phi_{2}-(\partial_{2}\bar\Phi_{2})\Phi_{2}
\right)
\left(
\bar\Phi_{3}\Phi_{3}
\right)
\right\rangle
\end{aligned}
\end{align}
can be calculated by expanding and performing all possible Wick contractions.
% where the subscripts on the right side denote the superspace location of the fields and derivatives.

As an example, one of the nine Wick contraction terms in the above expansion is
\begin{align}
\begin{aligned}
\left\langle\bar\Phi_{1}(\partial_{1}\Phi_{1})\bar\Phi_{2}(\partial_{2}\Phi_{2})\bar\Phi_{3}\Phi_{3}\right\rangle&=\Big\langle
\wick[offset=1em,sep=.5em]{
\c3{\bar\Phi_{1}}\c1{(\partial_{1}\Phi_{1})}\c1{\bar\Phi_{2}}\c2{(\partial_{2}\Phi_{2})}\c2{\bar\Phi_{3}}\c3{\Phi_{3}}
}
\Big\rangle+\Big\langle
\wick[offset=1em,sep=.5em]{
\c1{\bar\Phi_{1}}\c2{(\partial_{1}\Phi_{1})}\c3{\bar\Phi_{2}}\c1{(\partial_{2}\Phi_{2})}\c2{\bar\Phi_{3}}\c3{\Phi_{3}}
}
\Big\rangle\\
&=\left(
\frac{1}{y_{13}}
\right)
\left(\partial_{1}
\frac{1}{y_{21}}
\right)
\left(
\partial_{2}\frac{1}{y_{32}}
\right)+\left(
\partial_{2}\frac{1}{y_{12}}
\right)
\left(\partial_{1}
\frac{1}{y_{31}}
\right)
\left(
\frac{1}{y_{23}}
\right)
\end{aligned}
\end{align}

Computing and collecting all the terms, the Wick contractions give
as expected, a linear combination of all the linearly indepenedent parity-even invariants with homogeneity $\lambda_{1}^{2}\lambda_{2}^{2}$,
\begin{align}
\langle J_{1}J_{1}J_{0}\rangle=\frac{-2}{\bar x_{12}\bar x_{23}\bar x_{31}}\left[
Q_{1}Q_{2}\left(
1+\frac9{16}R'
\right)+P_{3}^{2}\left(
2+\frac{3}{8}R'
\right)+2iP_{3}\bar R_{3}
\right]
\end{align}

% Two-point function
%\section{Conservation constraint example}
%\newpage

\bibliographystyle{JHEP}

\bibliography{N2SCFTv2}

\end{document}